\documentclass[default]{sn-jnl}

\jyear{2021}%

\theoremstyle{thmstyleone}%
\theoremstyle{thmstyletwo}%

\theoremstyle{thmstylethree}%

\usepackage{graphicx}
\usepackage{textcomp}
\usepackage{xcolor}
\usepackage{subfig}
\usepackage{hyperref}
\usepackage{multirow}
\usepackage{comment}

\newcommand{\ie}{\textit{i.e.}}
\newcommand{\eg}{\textit{e.g.}}
\newcommand{\Eg}{\textit{E.g.}}
\newcommand{\etal}{\emph{et al.}}

\newcommand{\revision}[1]{\textcolor{black}{#1}}
\newcommand{\revisionsec}[1]{\textcolor{black}{#1}}

\begin{document}

\title[A Simple yet Effective Framework for Active Learning to Rank]{A Simple yet Effective Framework for Active Learning to Rank}


\author{\fnm{Qingzhong} \sur{Wang}}

\author{\fnm{Haifang} \sur{Li}}

\author{\fnm{Haoyi} \sur{Xiong}}

\author{\fnm{Wen} \sur{Wang}}
\author{\fnm{Jiang} \sur{Bian}}
\author{\fnm{Yu} \sur{Lu}}
\author{\fnm{Shuaiqiang} \sur{Wang}}
\author{\fnm{Zhicong} \sur{Cheng}}
\author{\fnm{Dejing} \sur{Dou}}
\author{\fnm{Dawei} \sur{Yin\footnote[2]{Dawei Yin is the corresponding author, please contact him via \url{yindawei@acm.org}.}}}

\affil{Baidu Inc., \orgaddress{\city{Beijing},  \country{China}}}




\abstract{
While China has become the biggest online market in the world with around 1 billion internet users, Baidu runs the world largest Chinese search engine serving more than hundreds of millions of daily active users and responding to billions of queries per day. To handle the diverse query requests from users at web-scale, Baidu has done tremendous efforts in understanding users' queries, retrieving relevant content from a pool of trillions of webpages, and ranking the most relevant webpages on the top of results. Among these components used in Baidu search, learning to rank (LTR) plays a critical role and we need to timely label an extremely large number of queries together with relevant webpages to train and update the online LTR models. To reduce the costs and time consumption of queries/webpages labeling, we study the problem of \emph{Active Learning to Rank} (\emph{\bf active LTR}) that selects unlabeled queries for annotation and training 
 in this work. Specifically, we first investigate the criterion--\emph{Ranking Entropy (RE)} characterizing the entropy of relevant webpages under a query produced by a sequence of online LTR models updated by different checkpoints, using a Query-By-Committee (QBC) method. Then, we explore a new criterion namely \emph{Prediction Variances (PV)} that measures the variance of prediction results for all relevant webpages under a query. Our empirical studies find that RE may favor low-frequency queries from the pool for labeling while PV prioritizes high-frequency queries more. Finally, we combine these two complementary criteria as the sample selection strategies for active learning. Extensive experiments with comparisons to baseline algorithms show that the proposed approach could train LTR models to achieve higher Discounted Cumulative Gain (\ie, the relative improvement $\Delta$DCG$_4$=1.38\%) with the same budgeted labeling efforts.
}


\keywords{search, information retrieval, learning to rank, active learning, query by committee}



\maketitle

\section{Introduction}\label{intro}

Baidu has established herself as  the world largest Chinese search engine who serves hundreds of millions of daily active users and handles billions of queries per day. Till now, Baidu has archived trillions of webpages for search. In addition to webpages and data resources, Baidu has invented a number of most advanced technologies for search, ranging from language models for content understanding~\cite{sun2019ernie,sun2020ernie}, domain-specific recommendation~\cite{huang2018learning,huang2018improving,huang2020multi,fan2021meta}, online query-Ads matching for sponsored search~\cite{fan2019mobius,yu2020combo}, and software/hardware co-designed infrastructures~\cite{ouyang2014sda,zhao2019aibox,ouyang2020baidu} for handling web-scale traffics of online search.


Generally, ranking the retrieved contents plays a critical role in a search engine, 
where learning to rank (LTR) is a standard workhorse. 
To achieve better ranking performance, we need to use a large amount of annotated data to train an LTR model. However, it is extremely expensive and time-consuming to label the ranks of relevant webpages for every query~\cite{settles2011theories}. To address this issue, active learning~\cite{settles2009active,huang2021semi} to pick up a small number of most informative queries and relevant webpages for labeling is requested. 

In this paper, inspired by uncertainty-based 
active learning methods, we present a simple yet effective approach to active learning for ranking. First, we investigate 
\emph{Ranking Entropy (RE)}, which characterizes the uncertainty of the ranking for every relevant webpage under a query using a Query-By-Committee (QBC) method~\cite{freund1997selective}. Intuitively, RE could discover queries with ranking uncertainty --- \ie, the predicted ranks of webpages in a query are indistinguishable using the LTR model. However RE is also biased in favor of the low-frequency queries, \ie, the queries are less searched by users, as there are no sufficient supervisory signals (\eg, click-throughs) to train LTR models for fine predictions. The bias to the low-frequency queries would not bring sufficient information gain to LTR training. To alleviate this problem, we study yet another criterion -- \emph{Prediction Variance (PV)}, which refers to the variances of rank prediction results among all relevant webpages for a query. Intuitively, we assume a query pairing to multiple webpages that have clearly distinguished orders of ranking as a query with \emph{high diversity}.
%
%
We further assume the variance of rank prediction results would faithfully characterize the variance of ground truth rank labels--\ie, the diversity of webpages in a query. We thus propose to use PV as a surrogate for the diversity of webpages in a query. Please refer to \emph{Section 3} for detailed comparisons and empirical analysis with real data.

More specifically, we report our practices in using the above two criteria to design query selection strategies for active learning. We conducted comprehensive empirical studies on these two criteria using realistic search data. The empirical studies show that the use of RE results in bias to the low-frequency queries, while the use of PV leads to the potential over-fittings to the high-frequency queries. When incorporating low-frequency queries (less searched queries) in labeling, the active learner might not be able to train LTR models well, due to the lack of supervisory signals (\eg, click-throughs) to distinguish the webpages for the queries. In contrast, when using high-frequency queries (hot queries) in labeling, the active learner might not be able to adapt the out-of-distribution queries (which is critical for ranking webpages at web-scale). Please see also in \emph{Sections 3.2 and 3.3} for the details of the criterion and empirical observations.

Finally, extensive experiments with comparisons to baseline algorithms show that the proposed approach (\ie, the combination of RE and PV) could train LTR models to achieve higher accuracy with fewer webpages labeled. Specifically, we have made contributions as follows.
\begin{itemize}
    \item We study the problem of active learning for ranking in Baidu search engine, where we focus on selecting queries together with relevant webpages for annotations to facilitate LTR models training and updates. And we deploy the system in Baidu search engine. 
    
    \item In the context of Baidu search, we first consider commonly-used uncertainty metrics for active learning of LTR, namely \emph{Ranking Entropy (RE)}. We find the use of RE could be biased by the frequency of queries, \ie, low-frequency queries normally have higher RE scores, as LTR models usually have not been well trained to rank webpages in such queries due to the lack of supervisory signals. To de-bias RE, we propose to study yet another diversity-based criterion -- \textit{Prediction Variance (PV)} that may favor high-frequency queries and are highly correlated to the true label variance of webpages under the query. In this way, we combine the two criteria for additional performance improvements.
    
    \item We conduct extensive experiments, showing that our proposed approach is able to significantly improve the performance of LTR in the context of Baidu search. Specifically, we compare our proposals (the combination of RE and PV) with a wide range of sample selection criteria for active learning, including random pickup, and expected loss prediction (aka ELO-DCG)~\cite{long2014active}. The comparisons show that our proposals outperform other criteria, which discovers 43\% more validated training pairs and improves 
    DCG (\eg, $\Delta$DCG$_4$=0.35\%$\sim$1.38\% in offline experiments and $\Delta$DCG$_4$ = 0.05\%$\sim$0.35\% in online experiments) using the same budgeted labeling efforts under fair comparisons.
\end{itemize}
Note that in this work, we focus on the low-complexity criteria of sample selection in active learning for LTR. There also exists some sample set selection algorithms~\cite{wei2015submodularity,hanneke2015minimax} for active learning in the high-order polynomial or even combinatorial complexity over the number of unlabeled samples, which is out of the scope of this paper as we intend to scale-up active learning of LTR with large-scale unlabeled queries and webpages.

\section{Related Works}\label{rw}

The goal of \textit{active learning} (AL) is to select the most informative samples in the unlabelled data pool for annotation to train a model~\cite{active1996}. Generally, AL models are able to achieve similar performance but use fewer annotated data points. To select the most informative samples for labeling, two categories of methods, \ie, diversity-aware criteria and uncertainty-aware criteria, for sample selection have been studied. The diversity-aware methods~\cite{diverse2004,diverse2010} measure the diversity of every subset of unlabeled samples and select the sample set with top diversity for labeling, where the core-set selection \cite{coreset} leveraging core-set distance of intermediate features is a representative method here. While diversity-aware methods work well on small datasets, they might fail to scale up over large datasets due to the need for subset comparisons and selections.

The uncertainty-aware methods \cite{kapoor2007active,erm2015,drop2016,drop2017,bayesian2019,zhan2021comparative,zhan2022comparative,zhan2022pareto} screen the pool of unlabeled samples and select samples with top uncertainty in the context of training model (\eg, LTR models here) for labeling. While uncertainty-aware methods could easily scale up over large datasets due to the low complexity, a wide variety of uncertainty criteria have been proposed, such as Monte Carlo estimation of expected error reduction \cite{montecarlo2001}, distance to the decision boundary \cite{boundary2001,boundary2003}, the margin between posterior probabilities \cite{margin2006}, and entropy of posterior probabilities \cite{entropy2008,entropy2009,entropy2013}. \revision{W. Cai \etal~\cite{cai2014active} propose an active learning method based on the maximum model change (MMC) and apply it to the classification task. Similarly, EMCM~\cite{cai2013maximizing} employs the expected model change maximization to estimate the uncertainty, and EMCM is applied to the regression task. B. Settles \etal~\cite{settles2007multiple} introduce active learning into multiple-instance learning and apply active learning to multiple-instance logistic regression. The selection criterion also depends on uncertainty, \ie, the model calculates multiple-instance uncertainty using the derivative of bag output with respect to instance output. Note that both MMC and EMCM use gradients to estimate the model change, however, computing gradients is normally more difficult for an LTR model than using query-by-committee (QBC) to estimate uncertainty. In addition, MMC and EMCM are designed for classification and regression tasks, while we focus on a more challenging task --- webpage ranking.} 

\paragraph{Discussion.}
The most relevant works to this study are~\cite{long2010active,bilgic2012active,long2014active,cai2015active}. As early as 2010, Long \emph{et al.}~\cite{long2010active,long2014active} proposed the expected loss optimization (ELO) framework, which selects and labels the most informative unlabelled samples for LTR and incorporates a predictor for discounted cumulative gain (ELO-DCG) to estimate the expected loss of given queries and documents. The work~\cite{bilgic2012active} further confirmed ELO with DCG could work well with any rankers at scale and deliver robust performance. Cai \emph{et al.}~\cite{cai2015active} followed the settings of ELO and extended DCG by incorporating the kernel density of queries, so as to balance sample distribution and the model-agnostic uncertainty for sample selection. Compared to the above studies, this work revisits the problem of active learning for LTR at web-scale in the 2020s, and we study new metrics of uncertainty for query selection with online LTR performances reported and data analyzed in the context of Baidu search.

\begin{figure}[t]
    \centering
    \includegraphics[width=0.8\linewidth]{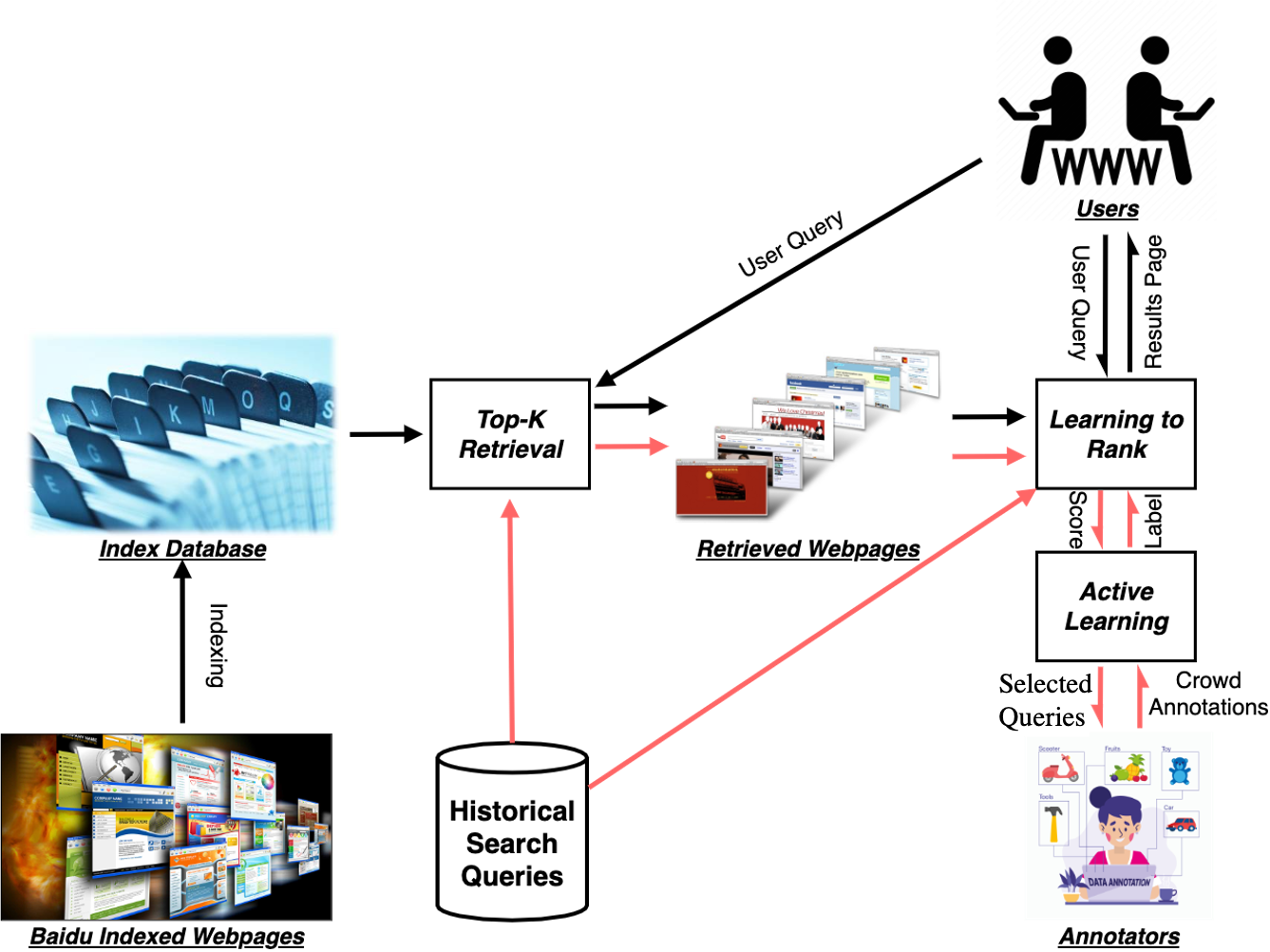}
    \caption{An overview of Baidu search system with the proposed active learning process. 
    While the search engine records every search query from users and stores them in \emph{Historical Search Queries}, it periodically picks up the NEW queries that appeared within the last ONE month for annotation and re-trains LTR models with annotated data.}
    \label{system}
\end{figure}

\section{Practical Active Learning to Rank for Web Search: Sample Selection Criteria and Empirical Studies}\label{method}
In this section, we first review the system design of \emph{active learning to rank} (Active LTR) for web search, then present our proposed selection criteria for active learning with empirical observations. 

\subsection{Active Learning to Rank (LTR) for Web Search at Baidu}
As shown in Figure~\ref{system}, given a search query, denoted as $q$, from a user, the search engine frequently first retrieves all relevant webpages, denoted as $\{w_1,w_2,\dots\}$, from the dataset and sorts the top-$K$ relevant webpages for the best user reading experience through ranking. To rank every webpage under the query, the search engine pairs every webpage with the query to form a query-webpage pair, e.g., $(q,w)$, and then extracts features from $(q,w)$, denoted as the feature vector $(x_q, x_w)$, where $x_q$ denotes query-relevant features and $x_w$ denotes webpage-relevant features and adopts the \emph{learning to rank} (LTR) model to predict the ranking score, e.g., \{\textbf{bad, fair, good, excellent, perfect}\}\footnote{For human annotations, labels 0, 1, 2, 3, 4, 5 denote bad, fair, good, excellent and perfect, respectively.} at Baidu Search using $(x_q, x_w)$.

To train the LTR model, the search engine usually collects the \emph{Historical Search Queries} $\mathcal{Q}=\{q_1,q_2,\cdots\}$ and archives relevant webpages $\mathcal{W}=\{w_{1,1}, w_{1,2}, \cdots\}$, where $w_{i,j}$ denotes the $j$th webpage associates with $q_i$. To scale up the \emph{active LTR} on trillions of webpages/queries while ensuring the timeliness of a search engine, our active learning system (\textcolor{red}{red} path in Figure~\ref{system}) periodically picks up NEW queries appeared within the last ONE month i.e., $\mathrm{S}\subset\mathcal{Q}$, pairs every selected query in $\mathrm{S}$ with retrieved webpages and extracts feature vectors to form the unlabeled datasets denoted as $\mathcal{T}=\{(x_{q_1},x_{w_{1,1}}), (x_{q_1}, x_{w_{1,2}}),\cdots\}$. Finally, the search engine recruits annotators to label $\mathcal{T}$ and retrains the LTR model with annotated data.



\subsection{Sample Selection Criteria for Active LTR}
In this section, we present the two criteria proposed for active learning to rank webpages.

\subsubsection{Ranking Entropy (RE)}
Uncertainty is one of the most popular criteria in active learning and \textit{Query-By-Committee} (QBC) \cite{freund1997selective} approach has been widely applied to estimate the uncertainty scores of the unlabelled data. In this paper, we apply QBC to compute the \textit{Ranking Entropy} (RE) of each webpage. Normally, there are $M$ models $\{h_{m}(x_{q}, x_{w}); m=1, \cdots, M\}$ to constitute a committee. Given the representation of a query-webpage pair $(x_{q_i},x_{w_{i,j}})\in\mathcal{T}$, the committee would provide a set of scores $\mathcal{S}_{i,j}=\{h_m(x_{q_{i}},x_{w_{i,j}}); m=1,\cdots, M\}$. Then, for any two webpages $\{w_{i,u},w_{i,v}\}$ associate to $q_i$, we can easily calculate the probability that webpages $w_{i,u}$ is ranked higher than $w_{i,v}$ under query $q_i$, denoted as the probability of $w_{i,u} \succ w_{i,v}$, \ie,
\begin{equation} \label{eq2}
    \pi_{i}^m (w_{i,u}\succ w_{i,v}) = \frac{1}{1+exp\left(\frac{-h_m(x_{q_i},x_{w_{i,u}}) + h_m(x_{q_i},x_{w_{i,v}})}{T}\right)},
\end{equation}
where $T$ denotes the temperature and $w_{i,u} \succ w_{i,v}$ denotes that $w_{i,u}$ is more relevant than $w_{i,v}$ under query $q_i$.

Similar to SoftRank \cite{taylor2008softrank}, we can obtain the distribution over ranks based on the probabilities calculated by Eq. (\ref{eq2}). We define the initial rank distribution of the $v$th webpage $w_{i,v}$ as $p_{i,v}^{1,m}(r) = \delta(r)$, where $\delta(r)=1$ if $r=0$ and 0 otherwise, then we can calculate the distribution in the $k$th step as follows:
\begin{equation}
\begin{aligned}
    p^{k,m}_{i,v}(r) = &p_{i,v}^{k-1, m}(r-1)\pi_i^m(w_{i,u}\succ w_{i,v}) + \\
    &p_{i, v}^{k-1, m}(r)\left(1-\pi_{i}^m(w_{i,u}\succ w_{i,v})\right),
\end{aligned}
\end{equation}
and the distribution in the last step will be the final ranking distribution. The computing procedure is shown in Alg. \ref{alg1}, where one webpage is added to the list for comparison in each step and the ranking distribution is updated using Eq. (\ref{eq2}).
  
For each webpage $w_{i,j}$, we can obtain a set of ranking distributions $\{p_{i,j}^m(r);m=1,\cdots, M;r=1, \cdots,N_i\}$ using Alg. \ref{alg1}, where $N_i$ denotes the number of webpages associate with $q_i$. Finally, for every query-webpage pair with the feature vector $(x_{q_i},x_{w_{i,j}})\in\mathcal{T}$, we use the average distribution over the committee to compute its entropy score as follow
\begin{align}
    p_{i,j}(r) &= \frac{1}{M}\sum_{m=1}^M p_{i,j}^m(r), \\
    E_{i,j} &= -\sum_{r=1}^{N_i} p_{i,j}(r)\log_2p_{i,j}(r).
\end{align}
Note that the goal of this paper is to select queries, hence, for a query $q_i$, we employ the average entropy of the webpages $\{w_{i,j};j=1,\cdots, N_i\}$ that associate with $q_i$, \ie, \begin{equation}
    RE(q_i) = \frac{1}{N_i}\sum_{j=1}^{N_i} E_{i,j}. 
\end{equation}
Higher $RE(q_i)$ refers to larger uncertainty in ranking results across the LTR models in the committee. Active learners are expected to pickup queries with large \emph{Ranking entropy}, i.e., higher $RE(q_i)$ for $q_i\in\mathcal{Q}$, for annotation and training. 

\begin{algorithm}[t]
\caption{Ranking Distribution}\label{alg1}
\begin{algorithmic}[1]
\Require{$\{\pi_i^m(w_{i,u}\succ w_{i,v});u,v=1,\cdots,N_i; u \neq v\}$}
\Ensure{$\{p_{i,v}^m;v=1,\cdots, N_i\}$}
\For {$v$ in range($N_i$)}
    \State $p_{i,v}^m=[1, 0, \cdots, 0]$ \Comment{$N_i$ elements}
    \State $p_{tmp} = [0,\cdots, 0]$ \Comment{$N_i$ elements}
    \State $\pi_v=\left[\pi_i^m(w_{i,v}\succ w_{i,1}),\cdots, \pi_i^m(w_{i,v}\succ w_{i,N_i})\right]$ 
    
    \For{u in range(1, $N_i$)}
        \For{r in range($v+1$)}
            \If{r==0}
                \State $\alpha=0$
            \Else
                \State$ \alpha=p_{i,v}^m[r-1]$
            \EndIf
            \State $p_{tmp}[r]=p_{i,v}^m[r] * \pi_{v}[u-1] + \alpha * (1- \pi_{v}[u-1])$
        \EndFor
        $p_{i,v}^m=p_{tmp}$
    \EndFor
\EndFor
\end{algorithmic}
\end{algorithm}

\subsubsection{Prediction Variance (PV)}
In our work, we assume a query, pairing to multiple webpages that are with clearly distinguished orders of ranking, as a \emph{query with high diversity}. While the true orders of ranking could be obtained through human annotations (\ie, labeling every webpage under the query using scores of five levels 0,~1,~2,~3,~and~4 in this work), we propose to use the rank prediction results of a trained LTR model to measure such diversity. Given a checkpoint of the online GBRank \cite{zheng2007regression} model, we propose to adopt the variance of predicted ranking scores (namely Prediction Variance (PV)
to measure the \emph{diversity of webpages} for the query.


Similar to RE, we also use the predictions of the committee to compute the prediction variance. Given the outputs of the committee $\mathcal{S}_{i,j}=\{h_m(x_{q_i},x_{w_{i,j}});m=1,\cdots, M\}$, the prediction variance of a query $q_i$ with $N_i$ retrieved webpages can be computed using the following equations:
\begin{align}
    \mu_m(q_i) &= \frac{1}{N_i}\sum_{j=1}^{N_i} h_m\left(x_{q_i},x_{w_{i,j}}\right),\\
    STD_m(q_i) &= \sqrt{\frac{1}{N_i}\sum_{j\in 1}^{N_i} \left(h_m\left(x_{q_i},x_{w_{i,j}}\right) - \mu_m(q_i)\right)^2},
\end{align}
Finally, we calculate the prediction variance $PV(q_i)$ as follows:
\begin{equation}
    PV(q_i) = \frac{1}{M}\sum_{m=1}^M STD_{m}(q_i).
\end{equation}
For active learning, we assume queries with large prediction variances, i.e., higher $PV(q_i)$ for $q_i\in\mathcal{Q}$, as the candidates for annotation and training.

\begin{figure*}[t]
  \centering
\subfloat[LV versus PV (Corr.: 0.59)]{\includegraphics[width=0.32\textwidth]{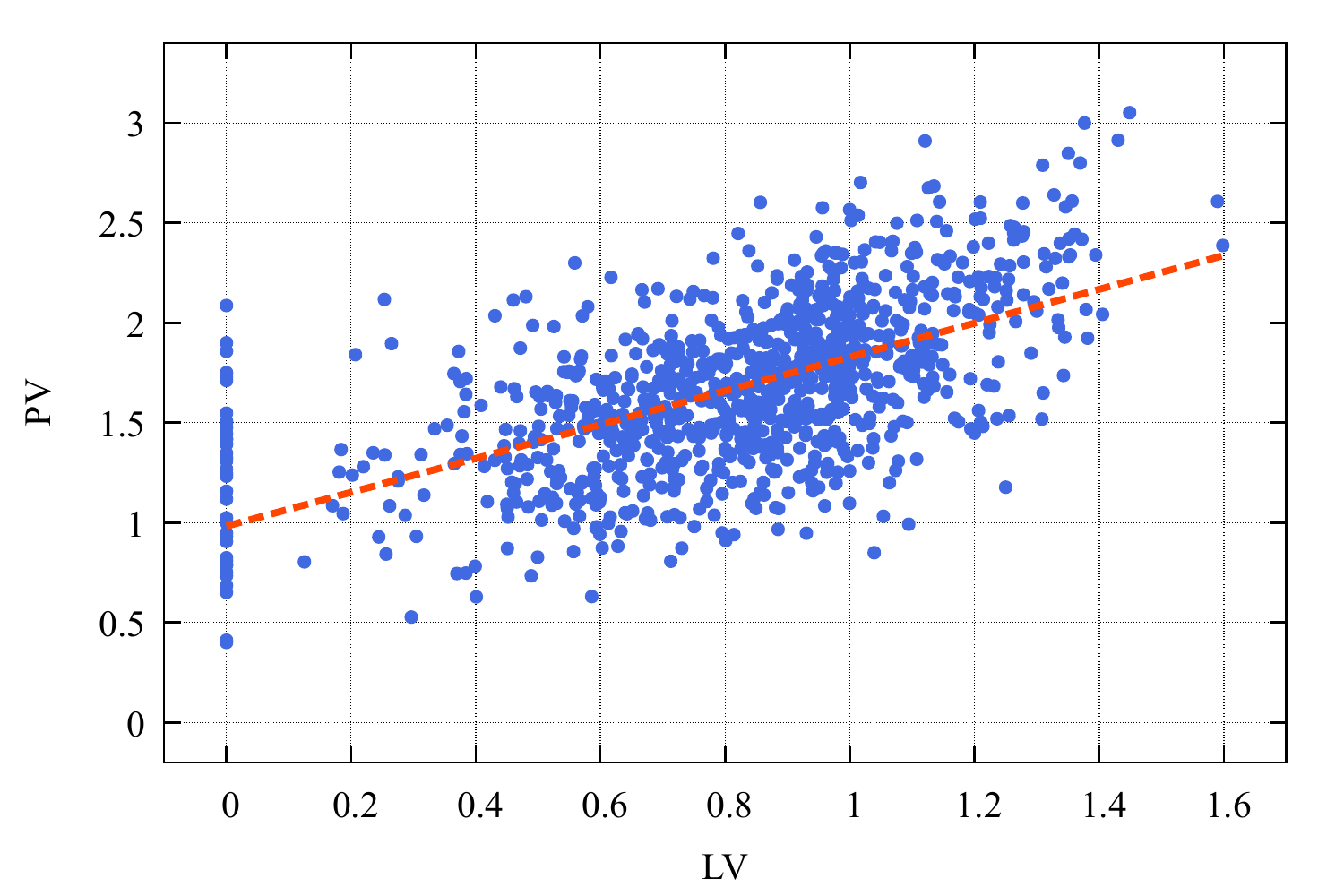}\label{lv-pv}}
\hfill
\subfloat[Best $DCG_4$ versus LV (Corr.: 0.66)]{\includegraphics[width=0.32\linewidth]{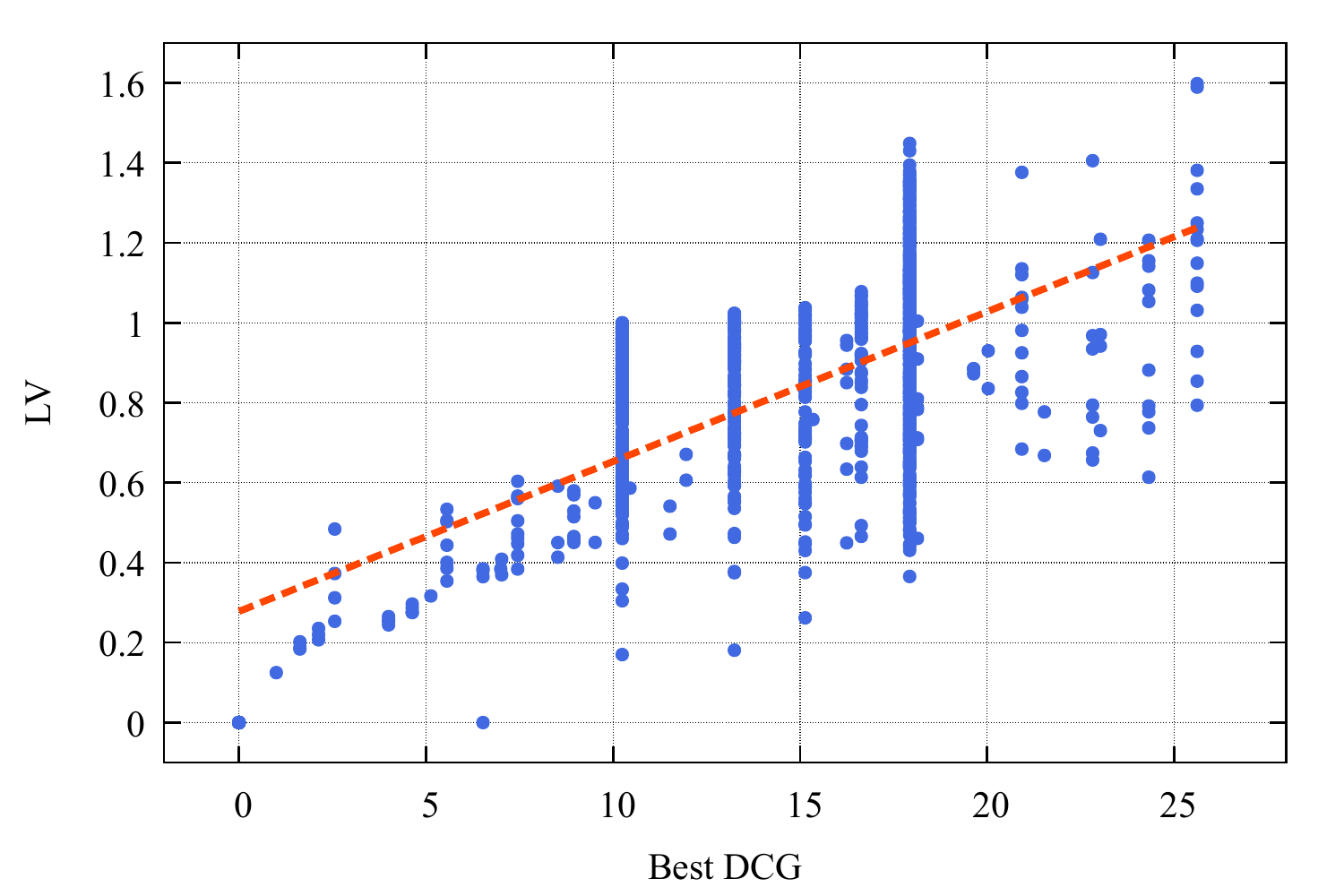}}
\hfill
\subfloat[Best $DCG_4$ versus PV (Corr.: 0.33)]{\includegraphics[width=0.32\textwidth]{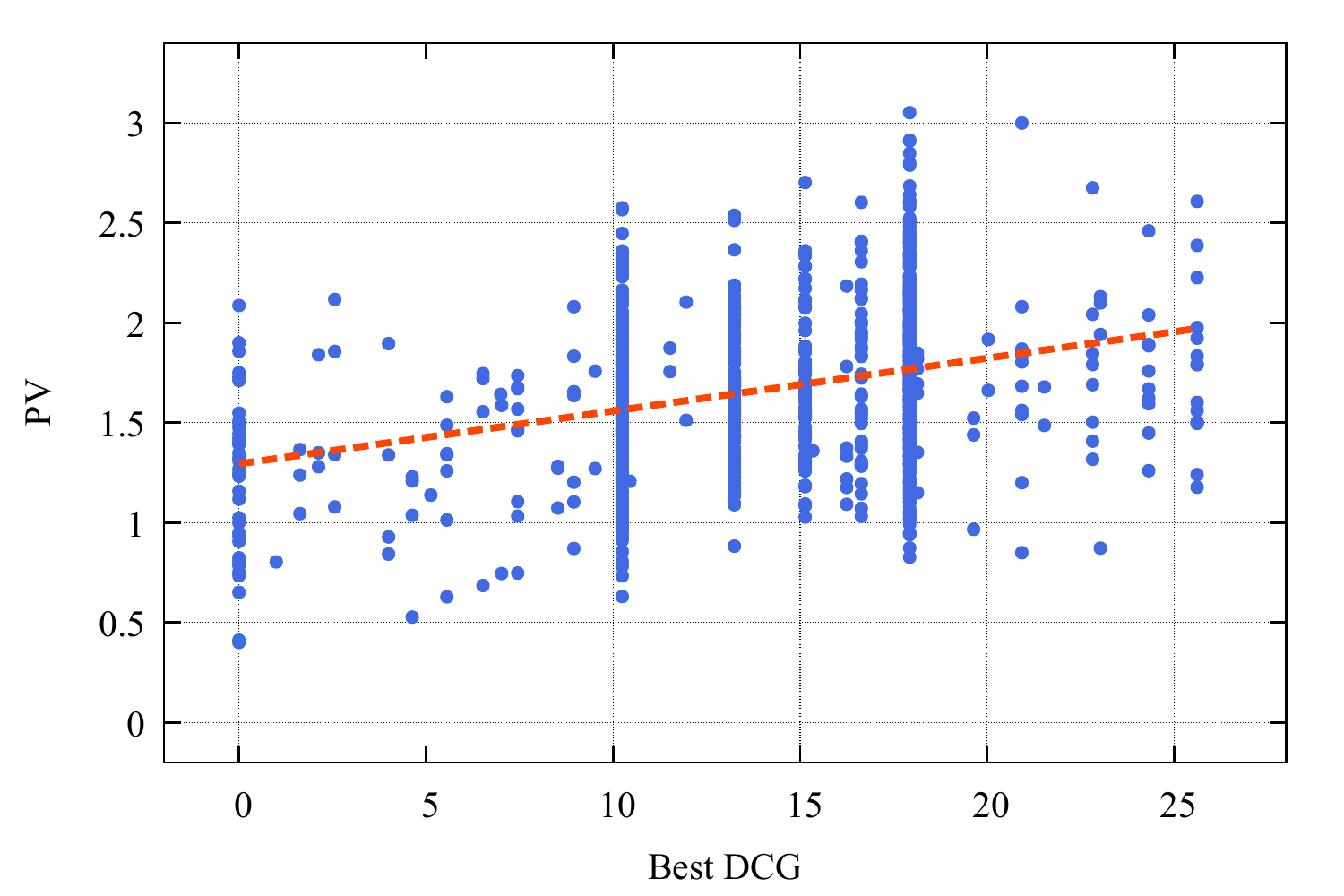}\label{dcg-re}}
\caption{Correlation Studies and Empirical Observations on Criteria based on 1000 Queries ($N=1000$). Best $DCG_4$ refer to the estimate of the upper bound of $DCG_4$, where we use the human-annotated ground truth labels to compute the $DCG_4$ score for every query.}
    \label{fig:empirical}
\end{figure*}

\subsection{Empirical Studies on Proposed Criteria for Active LTR}
While the first criterion \emph{Ranking Entropy (RE)} directly measures the uncertainty of ranking results under a query (either due to the defects of learned models or simply the difficulty to rank), we now hope to validate whether the second proposed criterion \emph{Prediction Variance (PV)} could characterize the diversity of webpages in a query and how PV improve LTR. We conduct empirical studies based on 1000 realistic query data drawn from the validation set and hope to test two hypotheses as follows.
\begin{itemize}
    \item \emph{Does PV characterize the variance of human-annotated ground truth ranking scores for LTR?} Here to test our hypothesis, we fetch a past checkpoint of the online LTR model in Baidu search and use the model to predict the ranking scores for every webpage in the 1000 queries. Note that, such 1000 queries are obtained from the validation set and have no overlapped with the queries used for training the LTR model. In Figure~\ref{fig:empirical}(a), we plot the scatter points of LV (label variance) versus PV with Pearson correlation $C\mathrm{orr}=0.59$ and $p$-value$<0.05$. In this way, we can conclude that PV significantly correlates to the variance of human-annotated ground truth ranking scores (LV) and faithfully characterizes the \emph{difficulty to rank} of every query.

    \item \emph{Does PV correlate to the information gain of query selection for LTR?} Yet another hypothesis in our mind is that selecting \emph{queries with a high diversity of webpages} for annotations could bring more information gain to LTR. We thus need to correlate LV and PV with certain information measures of queries. In this study, we use a measure namely Best $DCG_4$---referring to the estimate of the upper bound of $DCG_4$ that uses the human-annotated ground truth ranking scores as the prediction results of the ranking. Intuitively, the Best $DCG_4$ reflects the optimal $DCG_4$ that could be achieved by any algorithm. The correlation studies have been done and illustrated in Figures~\ref{fig:empirical}(b)~and~(c). The significance could be found in the correlations between LV and Best $DCG_4$ and the correlations between PV and Best $DCG_4$. The observations suggest that queries with higher \emph{diversity in webpages} usually are more informative for LTR, no matter if the diversity was measured by LV or PV.
\end{itemize}
Based on the above two observations, we could conclude that (1) PV could faithfully characterize LV (label variance -- the \emph{diversity of webpages}), though PV was estimated using the prediction results of a model, and (2) a query with higher LV or PV usually is more informative for LTR, since LTR models are normally trained using pairwise or list-wise loss and a small LV leads to small loss values and gradients.

\subsection{Combined Criteria}
To be simple, we use the weighted sum of $RE$ and $PV$ as the acquisition function in active learning. For each query $q_i\in\mathcal{Q}$, the acquisition function is
\begin{equation}\label{acquisition}
    f(q_i) = RE(q_i) + \alpha*PV(q_i),   
\end{equation}
where $\alpha$ is a hyper-parameter to balance ranking uncertainty and webpage variance. We select the queries that have the largest values of $f(q_i)$ in each cycle of active learning.


\begin{figure*}[t]
    \centering
    \includegraphics[width=\linewidth]{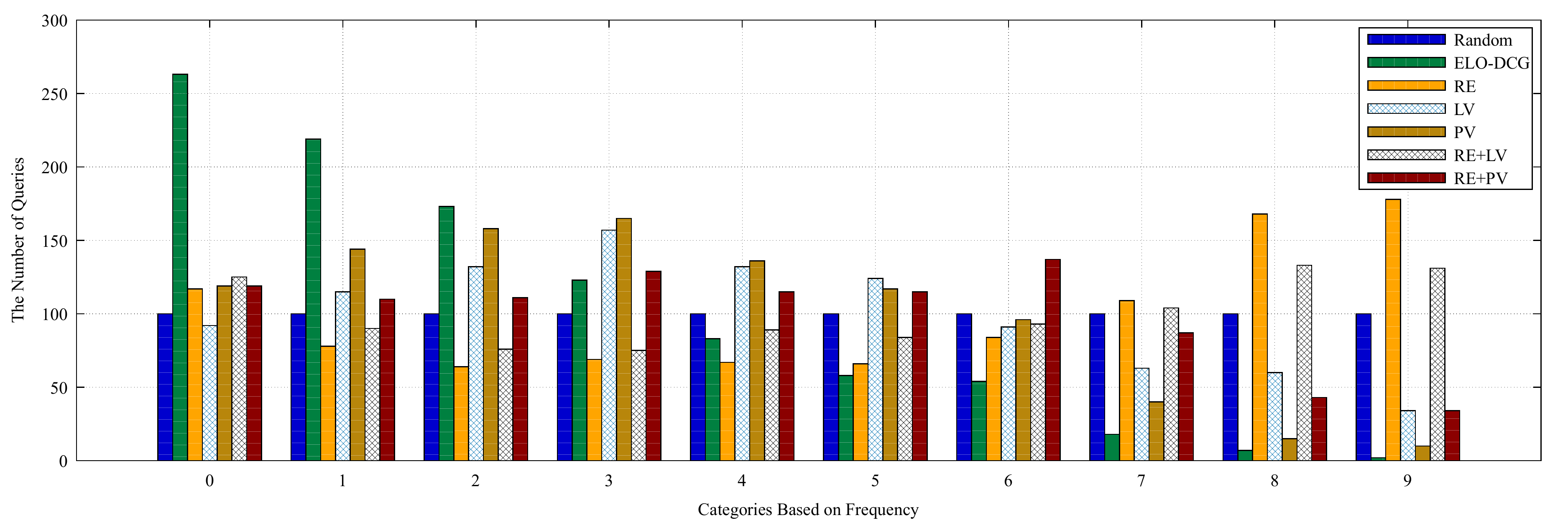}
    \caption{The distribution of 1,000 selected queries using different criteria. LV stands for label variance, PV for prediction variance and RE for ranking entropy.}
    \label{bucket_distribution}
\end{figure*}

\begin{figure*}[t]
    \centering
    \subfloat[Random]{\includegraphics[width=0.49\linewidth]{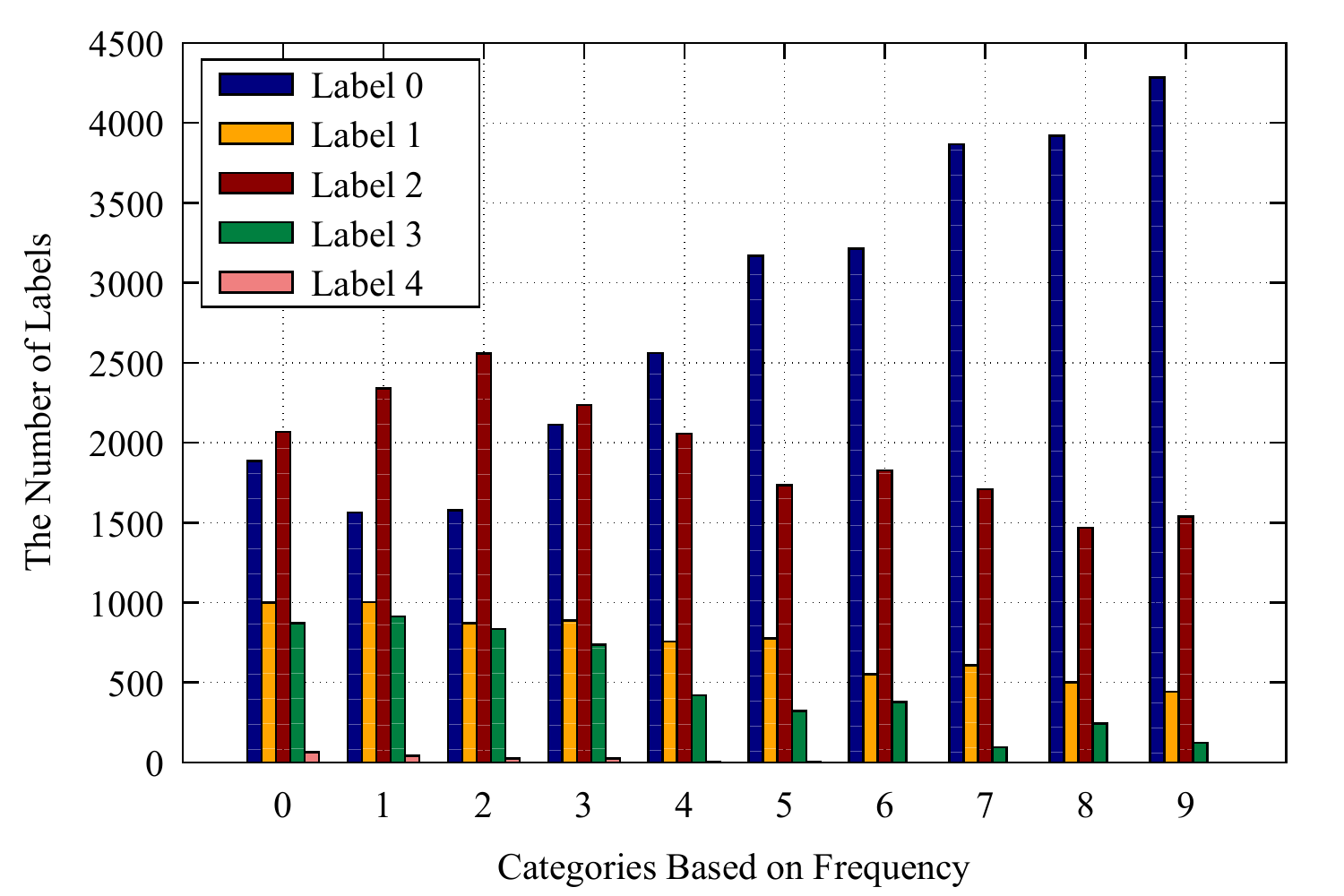}}\
    \hfill
    \subfloat[ELO-DCG]{\includegraphics[width=0.49\linewidth]{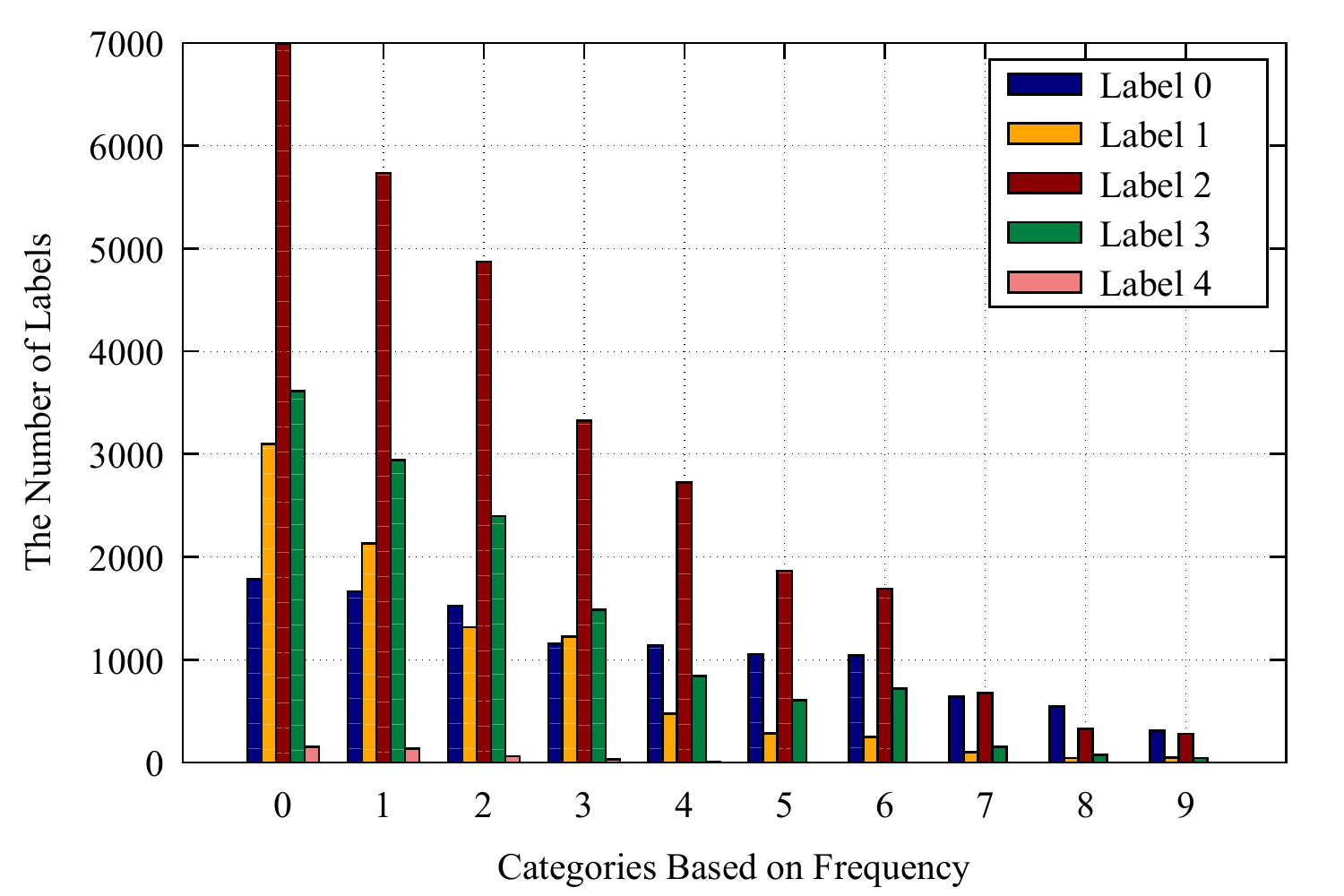}}\\
    \subfloat[RE]{\includegraphics[width=0.49\linewidth]{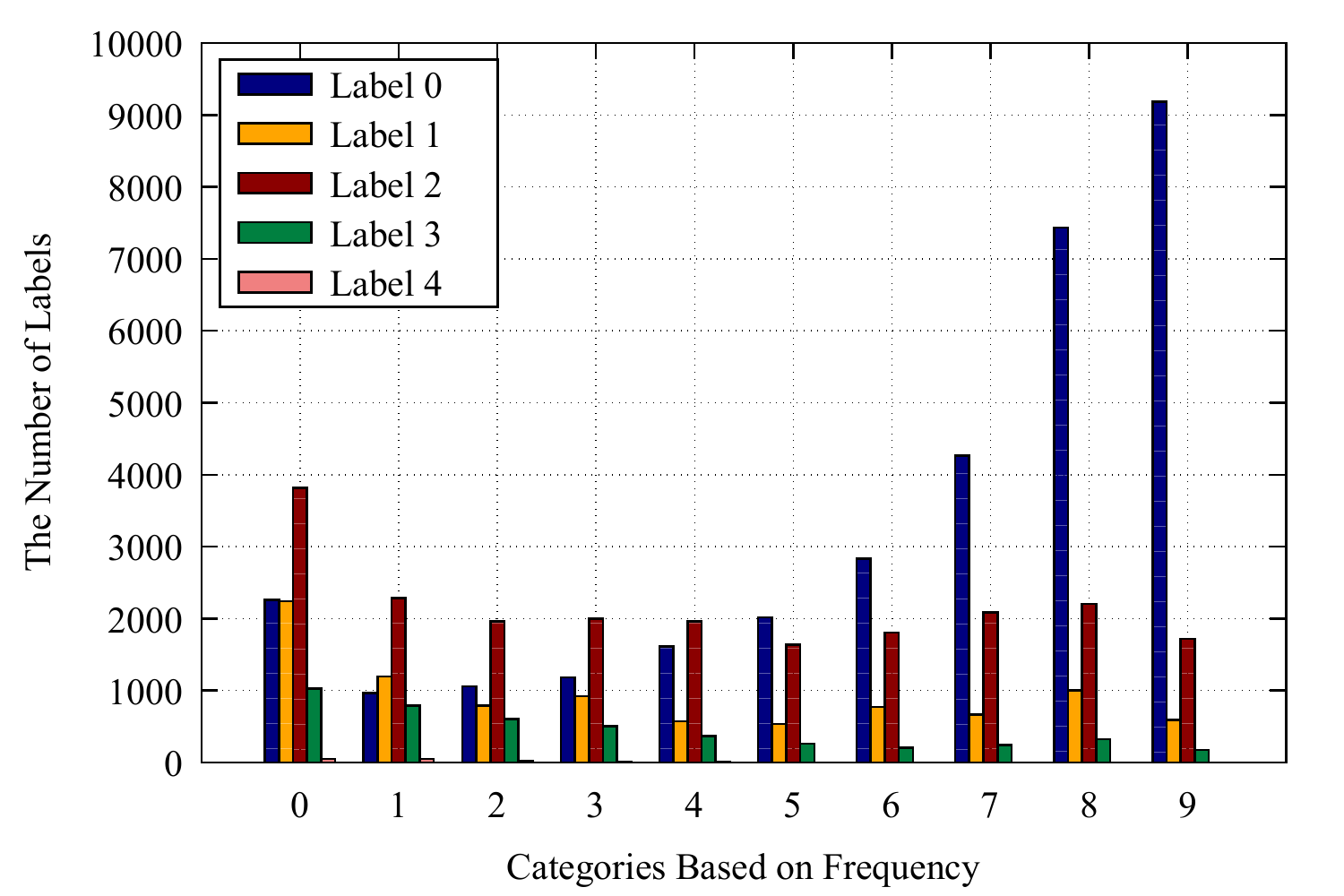}} \
    \hfill
    \subfloat[LV]{\includegraphics[width=0.49\linewidth]{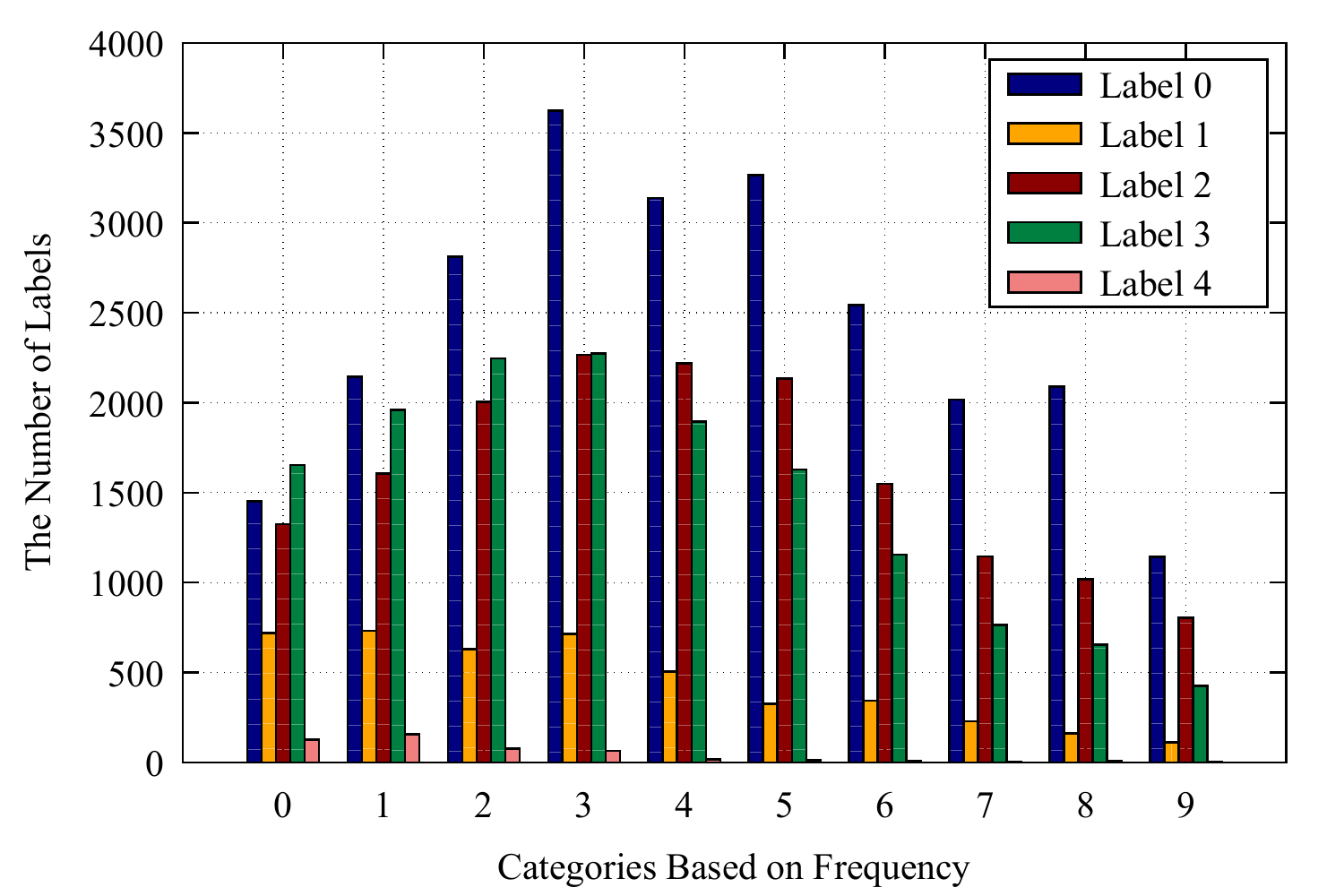}}\\
    \subfloat[PV]{\includegraphics[width=0.49\linewidth]{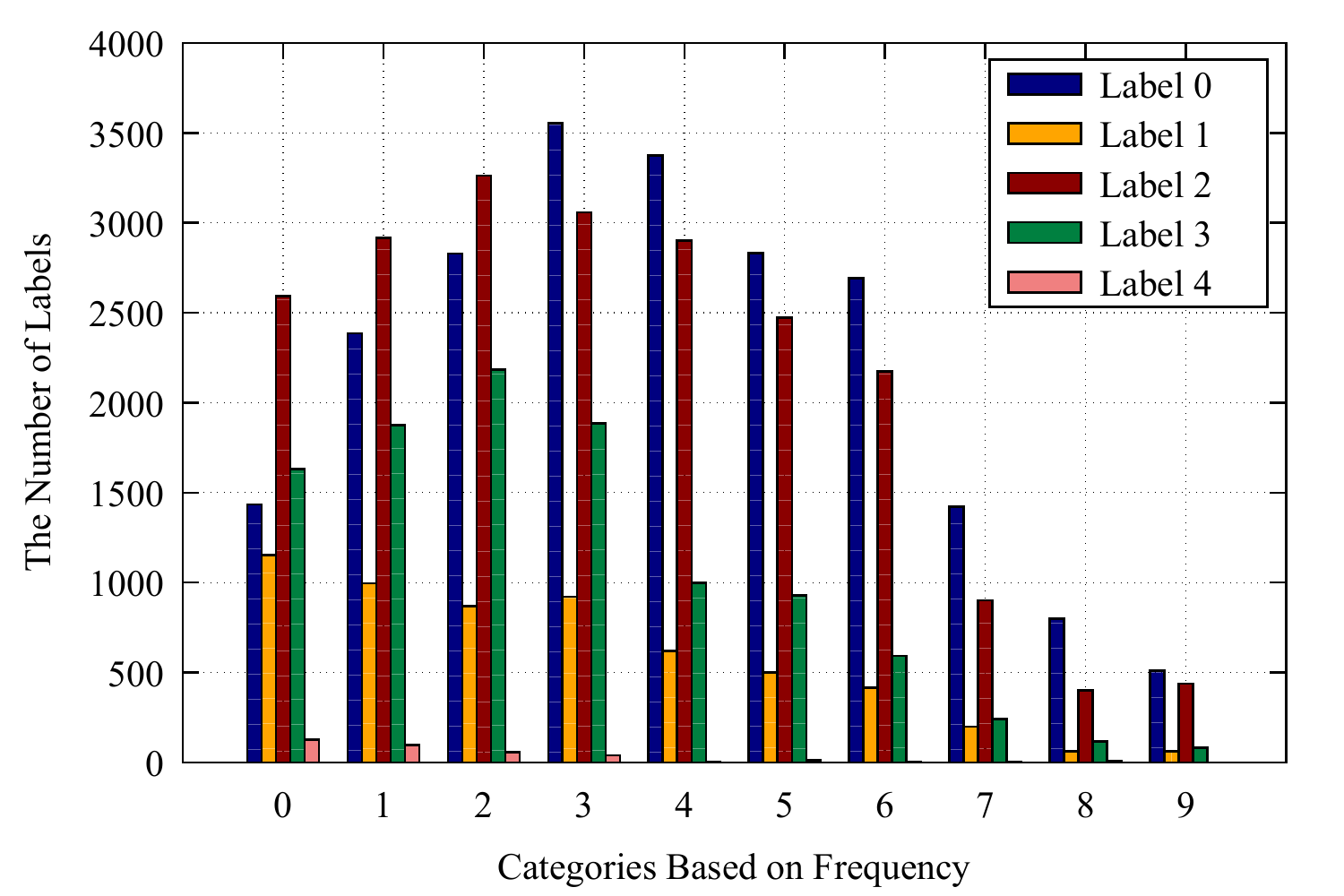}}\
    \hfill
    \subfloat[RE+PV]{\includegraphics[width=0.49\linewidth]{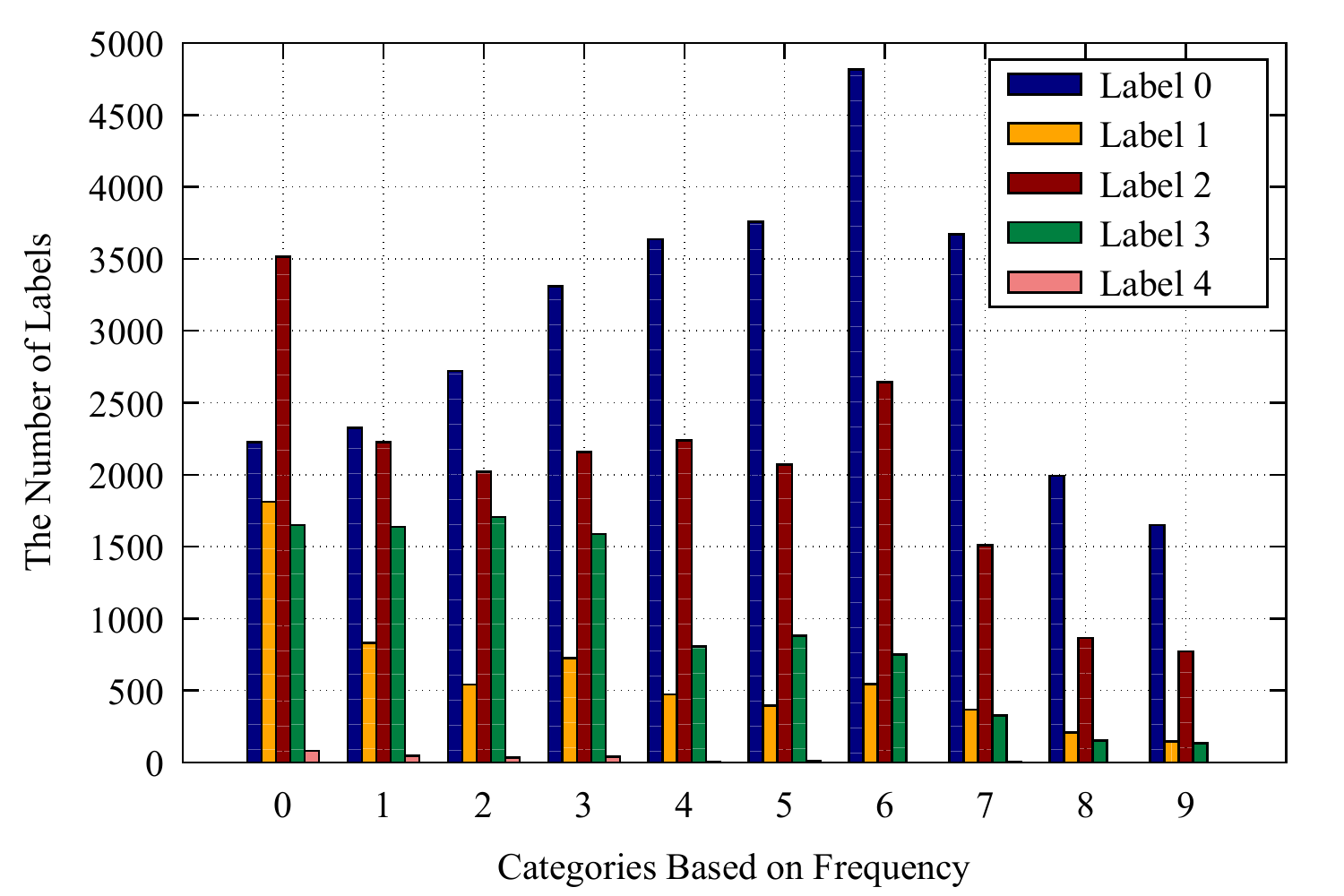}}
    \caption{The distribution of labels over categories. We use 1,000 selected queries and the corresponding webpages to obtain the statistics.}
    \label{label_distribution}
\end{figure*}

\begin{table}[t]
        \centering
        \begin{tabular}{c|c|c|c|c}
        \hline
            \multirow{2}{*}{Dataset} & \multicolumn{2}{c|}{Train} &\multicolumn{2}{c}{Validation \& Test} \\
            \cline{2-5}
            & \# queries & \# documents & \# queries & \# documents \\
            \hline
            Yahoo set1 \cite{chapelle2011yahoo} & 19,944 & 473,134 & 9,976 & 236,743 \\
            Yahoo set2 \cite{chapelle2011yahoo} & 1,266 & 34,815 & 5,064 & 138,005 \\
            Microsoft \cite{qin2010letor} & 18,900 & 2,261,000 & 12,600 & 1,509,000 \\
            Tiangong \cite{ai2018unbiased} & 3,449 & 333,813  & 100 & 10,000 \\
            Ours &14,000 & 840,000 & 1,000 & 60,000 \\
            \hline
        \end{tabular}
        \caption{\revision{The comparison between our dataset and existing datasets for learning to rank.}}
        \label{tab:dataset_comparison}
\end{table}

\section{Experiments}
In this section, we present the results of experiments, where we first introduce the results of offline experiments, then figure out the online performance of our proposals, both in comparisons with baseline algorithms.

\subsection{Offline Experiments and Results}
In this section, we present the details of offline experiments with introductions to the setups and results.

\subsubsection{Setups}

\revision{To conduct offline experiments, we construct a dataset for LTR. We classify the queries in the last month into 10 categories based on the frequency and then we filter out the erotic and illegal queries in each category. Finally, we randomly sample 1,500 queries from each category and for each query, we select 60 retrieved documents for human annotation, resulting in a dataset composed of 15,000 queries and 900,000 documents. Note that the dataset with 15,000 queries is relatively large and we present the comparison between our dataset and existing LTR datasets in Table \ref{tab:dataset_comparison}. In the dataset, the label of each query-document pair is in 5 levels: bad, fair, good, excellent and perfect and the corresponding relevant scores are \{0, 1, 2, 3, 4, 5\} respectively.}

\revision{To train the model, first, we split the dataset into a training set (14,000 queries) and a validation set (1,000 queries).} In the beginning of active learning, we randomly select $N_0$ queries from the training set as the base and in each cycle of active learning, we set the batch size $bs=100$, \ie, we select 100 queries from the pool (the rest of the training set) using the acquisition function \footnote{The reason for using $bs=100$ is that annotating the relevant scores are expensive and time-consuming. We can only annotate 500 queries per day and in the following offline experiments, we also consider $bs=500$.}. The quota is 2,000 queries, \ie, we run active learning for 20 cycles. We also conduct ablation studies on the value of $\alpha$ in Eq. (\ref{acquisition}), where $\alpha=\{0.5, 1.0, 1.5\}$. We set the number of committee $M$ to 9, \ie, 9 variants of GBRank with different numbers of trees (100, 300, 500) and maximum depth (1, 3, 5). 

To evaluate the performance of an LTR model, we use \textit{Discounted Cumulative Gain} (DCG), computing as follows:
\begin{equation}
    DCG_K = \sum_{k=1}^K \frac{G_k}{\log_2(k+1)},
\end{equation}
where $G_k$ denotes the weight assigned to the webpage's label at position $k$. A higher $G_k$ indicates that the webpage is more relevant to the query. Also, a higher $DCG_K$ indicates a better LTR model. In this paper, we consider the DCG of top 4 ranking results, \ie, $DCG_4$. In addition, we consider another important metric -- the percentage of the irrelevant webpages in top $K$, which is computed as follows:
\begin{equation}
    R_{01} = \frac{N_{01}}{K},
\end{equation}
where $N_{01}$ denotes the number of the irrelevant webpages\footnote{We consider the webpages with labels of 0 and 1 as irrelevant webpages.}. Obviously, a lower $R_{01}$ indicates a better LTR model. Also, we consider $R_{01}$ in the top 4 in this paper.

\revision{In addition to DCG and $R_{01}$, we also compare the distribution of the selected queries and the number of valid training pairs obtained by using different methods, which is able to reflect the label diversity of webpages. A large label diversity means that webpages are uniformly distributed on each label and a small diversity indicates that most webpages have the same label.
}

There are two baselines for comparison, the first one is random selection and the second one is ELO-DCG\cite{long2014active} --  an uncertainty-based active learning method for ranking.


\subsubsection{Offline results} Here, we first present the statistical characteristics of selected queries (with webpages retrieved) for annotations, then introduce the details about the valid query-webpage pairs formed from annotation results for training. Finally, we present the performance improvements of proposed criteria in comparison to baseline criteria, such as ELO-DCG~\cite{long2010active,long2014active}.

\paragraph{The distribution of selected queries}

Fig. \ref{bucket_distribution} shows the distribution over categories of 1,000 selected queries. Category 0 is composed of the most frequent queries, while category 9 contains the least frequent (only one time in the one-month search log) queries. For random selection, we randomly select 100 queries from each category. Compared with random selection, \textit{Label Variance} (LV) prefers relatively frequent queries, such as categories 2-4, but selects fewer low-frequency queries in categories 7-9, indicating that the webpages associated with low-frequency queries have similar human annotations. PV performs similarly in high-frequency queries but selects much fewer low-frequency queries. By contrast, RE selects the most low-frequency queries. However, it is difficult to construct enough training pairs if there are too many low-frequency queries since irrelevant webpages dominate these queries. In Fig. \ref{label_distribution}(c), we demonstrate the distribution of labels, where we use 1,000 randomly selected queries to obtain the statistics. Obviously, label 0 dominates low-frequency queries leading to difficulties on constructing training pairs for GBRank, hence, we need to balance the number of selected queries in each category and RE+PV is able to achieve the goal (see Fig. \ref{label_distribution}(f)). Interestingly, the existing work -- ELO-DCG \cite{long2010active,long2014active} is in favor of high-frequency queries. \revision{The reason is that for low-frequency queries, the best DCG and the DCG based on the average relevant scores are very small, leading to relatively small ELO-DCG.} Generally, Baidu search engine can well handle high-frequency queries to satisfy users' demands and selecting more high-frequency queries cannot benefit the gain of Baidu search. 

Looking at Fig. \ref{label_distribution}, where we present the distribution of labels over categories. For random selection, one observation is that low-frequency queries have more webpages with the label 0. In addition, the distributions of labels for each category are different and unbalanced. \revision{Using more webpages with label 0 is able to provide more training positive-negative pairs, for example, suppose we have $N$ relevant webpages and $M$ irrelevant webpages, then we can construct $N\times M$ training pairs and a larger $M$ enlarges the number of pairs, which could reduce the percentage of irrelevant webpages in top $K$ ranking results.} 
However, too many webpages with similar labels could reduce the number of valid training pairs, hurting ranking quality. By contrast with random selection, ELO-DCG \cite{long2010active,long2014active} selects more webpages with label 2 since it prefers high-frequency queries that have more relevant (label 2, 3, 4) webpages. Though ELO-DCG \cite{long2010active,long2014active} can select more relevant webpages, the diversity among webpages for each query is relatively low. While RE performs in the opposite way, selecting more webpages with label 0, which also lacks diversity among webpages. Intuitively, if the webpages of a query have the same label, then it is difficult to rank them, \ie, higher uncertainty. Moving on to Fig. \ref{label_distribution} (d-f) (LV, PV and RE+PV), the distribution of labels are more balanced, indicating the webpages are diverse. On one hand, a higher diversity score could result in more training pairs, on the other hand only considering diversity selects more queries that the trained model can easily rank the associated webpages and these queries cannot further improve the ranking model. On the contrary, RE+PV is able to balance diversity and uncertainty, selecting more useful and informative queries.

\paragraph{The number of training pairs}

\begin{table}[t]
    \centering
    \begin{tabular}{c|c|c}
    \hline
         Criterion & \# valid pairs & \# neg-pos pairs\\
         \hline
         Random &764,527 &534,500 \\
         ELO-DCG \cite{long2014active} &959,228 (25\%$\uparrow$) &598,555 (12\%$\uparrow$) \\
         LV &1,039,474 (36\%$\uparrow$) & 757492 (42\%$\uparrow$) \\
         PV &979,051 (28\%$\uparrow$) & 704,621 (32\%$\uparrow$) \\ 
         RE & 823,411 (8\%$\uparrow$) & 562,464 (5\%$\uparrow$) \\
         \textbf{Ours} (RE+PV) & \textbf{1,091,176 (43\%$\uparrow$)} & \textbf{803,723 (50\%$\uparrow$)}\\
         RE+LV(upper bound) & 1,149,083 (50\%$\uparrow$) & 827,133 (55\%$\uparrow$) \\
    \hline
    \end{tabular}
    \caption{The number of training pairs obtained by using different criteria and the relative improvement compared with random selection. If two webpages associated with a query have different labels, then they constitute a valid pair. Neg-pos pair denotes that a pair is composed of irrelevant and relevant webpages. We use each approach to select 1,000 queries to obtain the statistics.}
    \label{pairs}
\end{table}

\begin{figure*}[t]
    \centering
    \includegraphics[width=0.49\linewidth]{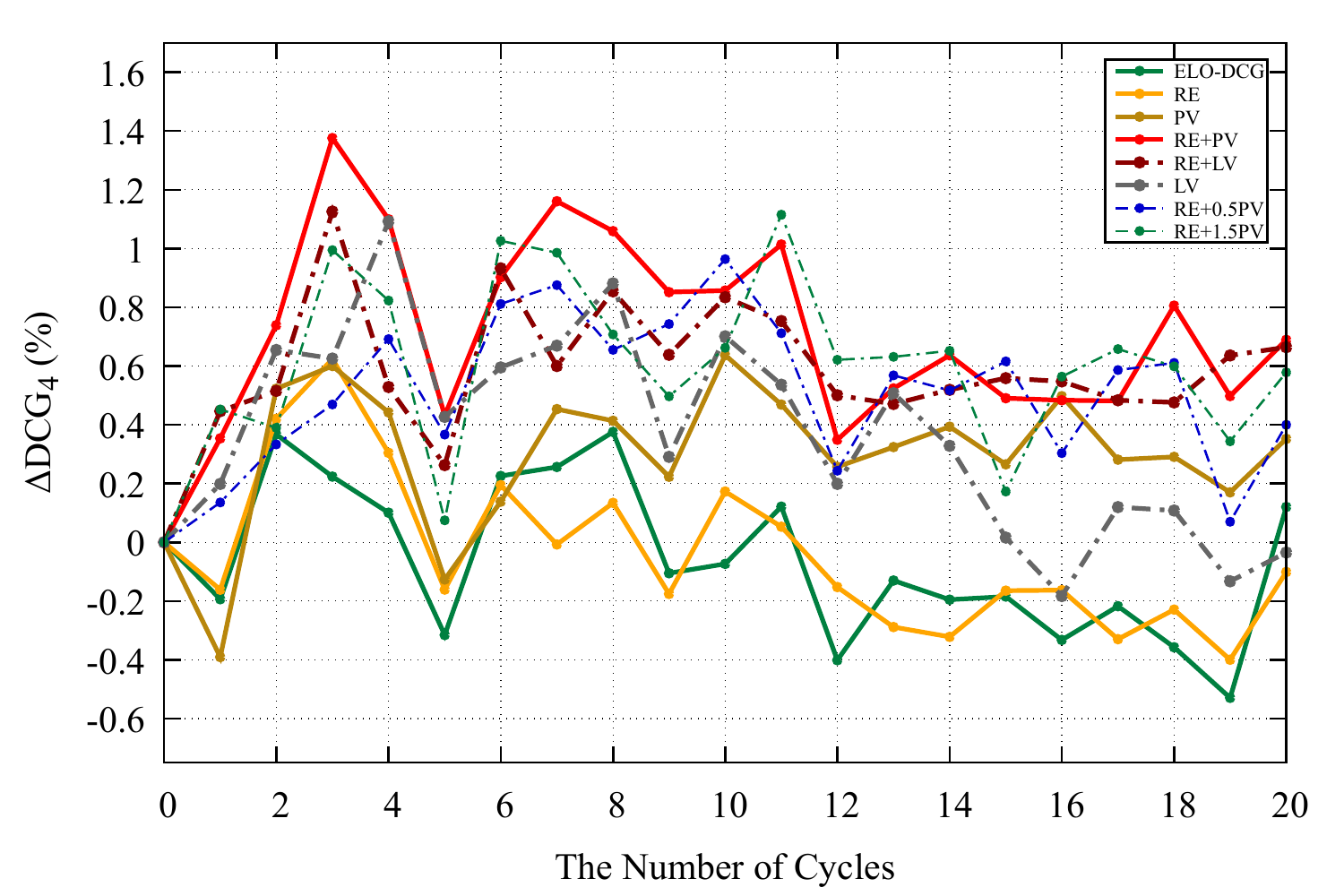}
    \hfill
    \includegraphics[width=0.49\linewidth]{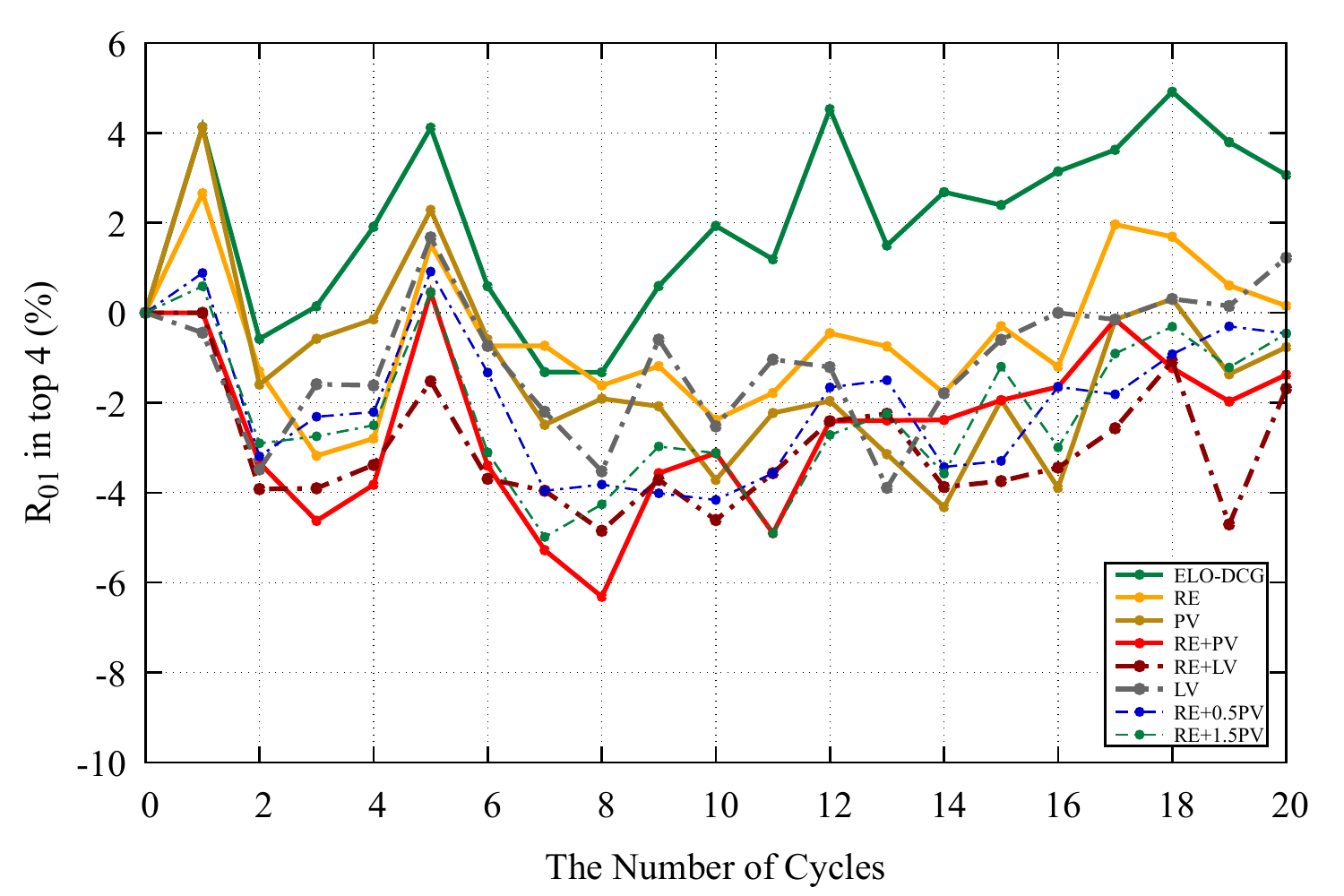} \\
    \includegraphics[width=0.49\linewidth]{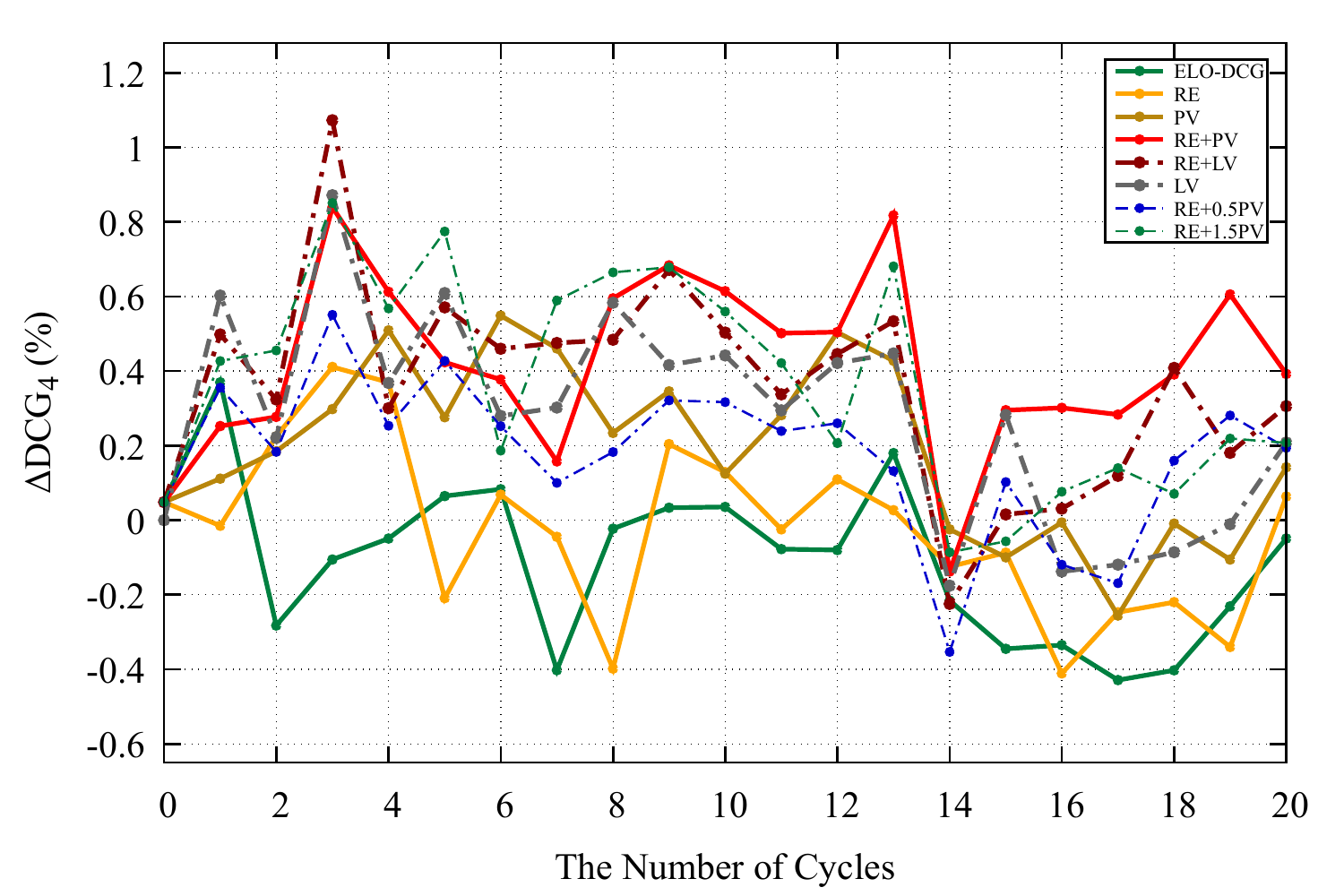}
    \hfill
    \includegraphics[width=0.49\linewidth]{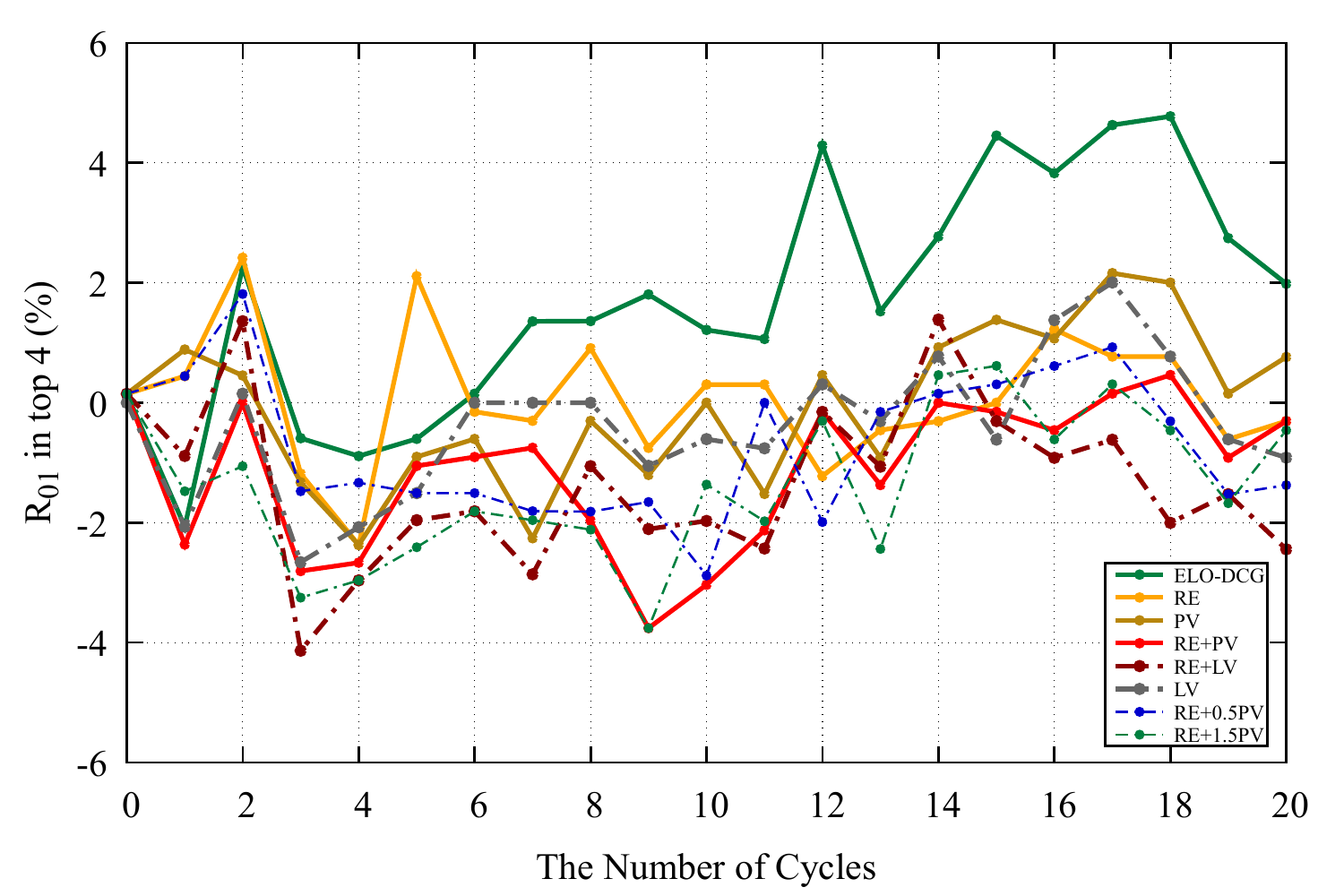}    
    \caption{
    The relative improvements of $DCG_4$ (\ie, $\Delta DCG_4$) and $R_{01}$ (\ie, $\Delta R_{01}$) compared with using random selection in each active learning cycle with the same budget. Top: the base set is composed of 100 queries. Bottom: the base set is composed of 500 queries.
    }
    \label{dcg-offline}
\end{figure*}

\begin{sidewaystable}[!htp]
    \centering
    \scalebox{0.9}{
    \begin{tabular}{c|c|cccccccccc|c}
    \hline
        \multirow{2}{*}{base} &\multirow{2}{*}{Method} & \multicolumn{10}{c|}{Categories based on frequency} & \multirow{2}{*}{All} \\
        \cline{3-12}
        &&0 &1 &2 &3 &4 &5 &6 &7 &8 &9 \\
        \hline
        \multirow{9}{*}{100}
        &ELO-DCG &
        \textcolor{red}{0.42\%$\uparrow$} 
        &
        \textcolor{blue}{0.14\%$\downarrow$} 
        &
        \textcolor{blue}{0.30\%$\downarrow$}
        &
        \textcolor{red}{0.31\%$\uparrow$}
        &
        \textcolor{red}{0.93\%$\uparrow$} 
        &
        \textcolor{blue}{0.18\%$\downarrow$} 
        &
        \textcolor{blue}{0.18\%$\downarrow$} 
        &
        \textcolor{red}{0.00\%$\uparrow$} 
        &
        \textcolor{blue}{1.44\%$\downarrow$} 
        &
        \textcolor{blue}{0.74\%$\downarrow$} 
        &
        \textcolor{red}{0.00\%$\uparrow$}\\
        &LV &
        \textcolor{red}{0.28\%$\uparrow$} 
        &
        \textcolor{blue}{0.29\%$\downarrow$} 
        &
        \textcolor{red}{0.22\%$\uparrow$} 
        &
        \textcolor{red}{0.78\%$\uparrow$} 
        &
        \textcolor{red}{0.51\%$\uparrow$} 
        &
        \textcolor{red}{0.72\%$\uparrow$} 
        &
        \textcolor{blue}{0.27\%$\downarrow$} 
        &
        \textcolor{red}{0.66\%$\uparrow$} 
        &
        \textcolor{blue}{0.24\%$\downarrow$} 
        &
        \textcolor{red}{0.25\%$\uparrow$} 
        &
        \textcolor{red}{0.43\%$\uparrow$}\\
        &PV &
        \textcolor{red}{0.50\%$\uparrow$} 
        &
        \textcolor{red}{0.07\%$\uparrow$} 
        &
        \textcolor{red}{0.15\%$\uparrow$} 
        &
        \textcolor{blue}{0.08\%$\downarrow$} 
        &
        \textcolor{red}{0.42\%$\uparrow$} 
        &
        \textcolor{red}{0.27\%$\uparrow$} 
        &
        \textcolor{red}{0.36\%$\uparrow$} 
        &
        \textcolor{red}{0.66\%$\uparrow$} 
        &
        \textcolor{red}{0.12\%$\uparrow$} 
        &
        \textcolor{blue}{0.62\%$\downarrow$} 
        &
        \textcolor{red}{0.35\%$\uparrow$}\\
        &RE &
        \textbf{1.13\%$\uparrow$} 
        &
        \textcolor{blue}{0.57\%$\downarrow$} 
        &
        \textcolor{red}{0.22\%$\uparrow$} 
        &
        \textcolor{blue}{0.54\%$\downarrow$} 
        &
        \textcolor{blue}{0.08\%$\downarrow$} 
        &
        \textcolor{red}{0.54\%$\uparrow$} 
        &
        \textcolor{red}{0.00\%$\uparrow$} 
        &
        \textcolor{blue}{0.28\%$\downarrow$} 
        &
        \textcolor{blue}{1.08\%$\downarrow$} 
        &
        \textcolor{red}{0.00\%$\uparrow$} 
        &
        \textcolor{red}{0.00\%$\uparrow$}\\
        &RE+0.5PV &
        \textbf{1.13\%$\uparrow$} 
        &
        \textcolor{red}{0.29\%$\uparrow$} 
        &
        \textcolor{red}{0.15\%$\uparrow$} 
        &
        \textcolor{red}{0.86\%$\uparrow$} 
        &
        \textcolor{red}{0.76\%$\uparrow$} 
        &
        \textcolor{red}{0.45\%$\uparrow$} 
        &
        \textcolor{red}{0.36\%$\uparrow$} 
        &
        \textcolor{red}{0.85\%$\uparrow$} 
        &
        \textcolor{blue}{0.48\%$\downarrow$} 
        &
        \textcolor{red}{0.62\%$\uparrow$} 
        &
        \textcolor{red}{0.52\%$\uparrow$}\\
        &RE+PV &
        \textcolor{red}{0.57\%$\uparrow$} 
        &
        \textbf{0.43\%$\uparrow$} 
        &
        \textcolor{red}{0.00\%$\uparrow$} 
        &
        \textbf{0.86\%$\uparrow$} 
        &
        \textcolor{red}{1.18\%$\uparrow$} 
        &
        \textcolor{red}{0.36\%$\uparrow$} 
        &
        \textbf{0.82\%$\uparrow$}
        &
        \textcolor{red}{1.14\%$\uparrow$} 
        &
        \textbf{0.96\%$\uparrow$}
        &
        \textbf{1.23\%$\uparrow$} 
        &
        \textbf{0.78\%$\uparrow$}\\
        &RE+1.5PV &
        \textcolor{red}{0.35\%$\uparrow$} 
        &
        \textcolor{red}{0.22\%} 
        &
        \textcolor{red}{0.30\%$\uparrow$} 
        &
        \textcolor{red}{0.70\%$\uparrow$} 
        &
        \textbf{1.43\%$\uparrow$} 
        &
        \textcolor{red}{0.36\%$\uparrow$} 
        &
        \textcolor{red}{0.45\%$\uparrow$} 
        &
        \textbf{1.42\%$\uparrow$}
        &
        \textcolor{red}{0.60\%$\uparrow$} 
        &
        \textcolor{red}{0.62\%$\uparrow$} 
        &
        \textcolor{red}{0.61\%$\uparrow$}\\
        &RE+LV &
        \textbf{1.13\%$\uparrow$}
        &
        \textcolor{red}{0.29\%$\uparrow$} 
        &
        \textbf{0.45\%$\uparrow$} 
        &
        \textcolor{red}{0.23\%$\uparrow$} 
        &
        \textcolor{red}{1.35\%$\uparrow$} 
        &
        \textbf{1.08\%$\uparrow$}
        &
        \textcolor{red}{0.64\%$\uparrow$} 
        &
        \textcolor{red}{0.85\%$\uparrow$} 
        &
        \textcolor{blue}{0.60\%$\downarrow$} 
        &
        \textcolor{red}{0.00\%$\uparrow$} 
        &
        \textcolor{red}{0.61\%$\uparrow$}\\
        \hline\hline
        \multirow{9}{*}{500}
        &ELO-DCG 
        &
        \textcolor{red}{0.28\%$\uparrow$}
        &
        \textcolor{red}{0.00\%$\uparrow$} 
        &
        \textcolor{red}{0.52\%$\uparrow$} 
        &
        \textcolor{blue}{0.39\%$\downarrow$}
        &
        \textcolor{blue}{0.17\%$\downarrow$}
        &
        \textcolor{blue}{0.36\%$\downarrow$}
        &
        \textcolor{blue}{0.54\%$\downarrow$}
        &
        \textcolor{blue}{0.28\%$\downarrow$}
        &
        \textcolor{blue}{0.48\%$\downarrow$} 
        &
        \textcolor{red}{0.00\%$\uparrow$} 
        &
        \textcolor{blue}{0.09\%$\downarrow$} \\
        &LV 
        &
        \textcolor{red}{0.08\%$\uparrow$}
        &
        \textcolor{blue}{0.21\%$\downarrow$}
        &
        \textbf{0.90\%$\uparrow$}
        &
        \textcolor{blue}{0.08\%$\downarrow$}
        &
        \textcolor{red}{0.17\%$\uparrow$} 
        &
        \textcolor{red}{0.91\%$\uparrow$} 
        &
        \textcolor{blue}{0.18\%$\downarrow$}
        &
        \textcolor{red}{0.19\%$\uparrow$}
        &
        \textcolor{red}{0.12\%$\uparrow$} 
        &
        \textcolor{red}{1.12\%$\uparrow$}
        &
        \textcolor{red}{0.26\%$\uparrow$} \\
        &PV 
        &
        \textcolor{blue}{0.14\%$\downarrow$} 
        &
        \textcolor{red}{0.00\%$\uparrow$}
        &
        \textcolor{red}{0.82\%$\uparrow$} 
        &
        \textcolor{blue}{0.77\%$\downarrow$}
        &
        \textcolor{red}{0.17\%$\uparrow$}
        &
        \textcolor{red}{0.63\%$\uparrow$} 
        &
        \textcolor{red}{0.00\%$\uparrow$} 
        &
        \textcolor{red}{0.19\%$\uparrow$} 
        &
        \textcolor{red}{0.48\%$\uparrow$} 
        &
        \textcolor{red}{0.99\%$\uparrow$} 
        &
        \textcolor{red}{0.17\%$\uparrow$} \\
        &RE 
        &
        \textcolor{red}{0.28\%$\uparrow$} 
        &
        \textcolor{blue}{0.29\%$\downarrow$} 
        &
        \textcolor{red}{0.37\%$\uparrow$} 
        &
        \textcolor{blue}{0.69\%$\downarrow$} 
        &
        \textcolor{blue}{0.75\%$\downarrow$} 
        &
        \textcolor{red}{0.36\%$\uparrow$}
        &
        \textcolor{red}{0.27\%$\uparrow$} 
        &
        \textcolor{blue}{0.17\%$\downarrow$} 
        &
        \textcolor{blue}{0.84\%$\downarrow$}  
        &
        \textcolor{red}{1.61\%$\uparrow$}  
        &
        \textcolor{blue}{0.09\%$\downarrow$}  \\
        &RE+0.5PV 
        &
        \textcolor{red}{0.35\%$\uparrow$} 
        &
        \textcolor{red}{0.29\%$\uparrow$}  
        &
        \textcolor{red}{0.75\%$\uparrow$}  
        &
        \textcolor{blue}{0.46\%$\downarrow$}  
        &
        \textcolor{blue}{0.42\%$\downarrow$} 
        &
        \textcolor{red}{0.63\%$\uparrow$}  
        &
        \textcolor{red}{0.09\%$\uparrow$}  
        &
        \textcolor{red}{0.09\%$\uparrow$}  
        &
        \textcolor{blue}{0.84\%$\downarrow$} 
        &
        \textcolor{red}{1.24\%$\uparrow$}  
        &
        \textcolor{red}{0.17\%$\uparrow$}  \\
        &RE+PV 
        &
        \textcolor{red}{0.07\%$\uparrow$}  
        &
        \textbf{0.43\%$\uparrow$} 
        &
        \textcolor{red}{0.22\%$\uparrow$}  
        &
        \textbf{0.23\%$\uparrow$}  
        &
        \textcolor{red}{0.17\%$\uparrow$}  
        &
        \textcolor{red}{0.27\%$\uparrow$}  
        &
        \textbf{0.63\%$\uparrow$} 
        &
        \textbf{0.47\%$\uparrow$} 
        &
        \textbf{1.20\%$\uparrow$}
        &
        \textcolor{red}{1.12\%$\uparrow$}  
        &
        \textbf{0.43\%$\uparrow$} \\ 
        &RE+1.5PV &
        \textbf{0.49\%$\uparrow$}
        &
        \textbf{0.43\%$\uparrow$}
        &
        \textcolor{red}{0.52\%$\uparrow$} 
        &
        \textcolor{blue}{0.15\%$\downarrow$}
        &
        \textbf{0.25\%$\uparrow$}
        &
        \textbf{0.72\%$\uparrow$} 
        &
        \textcolor{red}{0.36\%$\uparrow$} 
        &
        \textcolor{red}{0.28\%$\uparrow$} 
        &
        \textcolor{blue}{0.24\%$\downarrow$} 
        &
        \textcolor{red}{0.74\%$\uparrow$} 
        &
        \textcolor{red}{0.34\%$\uparrow$} \\
        &RE+LV 
        &
        \textcolor{red}{0.35\%$\uparrow$} 
        &
        \textbf{0.43\%$\uparrow$} 
        &
        \textcolor{red}{0.67\%$\uparrow$}
        &
        \textcolor{blue}{0.39\%$\downarrow$}  
        &
        \textcolor{red}{0.00\%$\uparrow$} 
        &
        \textbf{0.72\%$\uparrow$}
        &
        \textcolor{red}{0.36\%$\uparrow$} 
        &
        \textcolor{red}{0.28\%$\uparrow$}
        &
        \textcolor{blue}{0.12\%$\downarrow$}
        &
        \textbf{1.36\%$\uparrow$} 
        &
        \textcolor{red}{0.34\%$\uparrow$} \\
    \hline
    \end{tabular}
    }
    \caption{The relative improvement of average $DCG_4$ (\ie, $\Delta DCG_4$) over all active learning cycles compared with random selection. The \textcolor{red}{red} numbers represent the relative increases, the \textcolor{blue}{blue} numbers are the relative decreases and the \textbf{bold} numbers are the highest relative improvements. 
    }
    \label{avg_dcg}
\end{sidewaystable}

For GBRank \cite{zheng2007regression} that uses pairwise loss, the number of training pairs is crucial. With more training pairs fed to the training procedure, LTR models are expected to deliver better performance. Table \ref{pairs} presents the number of pairs obtained using different approaches. We can easily conclude that the proposed approach is able to obtain more training pairs compared with random selection and the existing work -- ELO-DCG \cite{long2010active,long2014active}. In terms of the number of valid pairs composed of two webpages with different human annotated labels, random selection obtains 764,527 pairs for 1,000 queries, while the number increases by 25\% when using ELO-DCG \cite{long2010active,long2014active}. Using the criteria LV and PV that improve the diversity among webpages, the number of valid pairs can be improved by 36\% and 28\%, respectively. Only using RE can also achieve a 8\% increase compared with random selection, but it is inferior to ELO-DCG, LV and PV since RE select more low-frequency queries and most webpages associated with these queries are with label 0 (see Fig. \ref{bucket_distribution} and Fig. \ref{label_distribution}(c)). Though selecting more low-frequency queries could benefit in solving the problem of unusual queries and attracting more users, it is difficult to retrieve relevant webpages for these queries and irrelevant webpages are less useful to train GBRank. Combining RE and PV is able to alleviate the problem. The number of valid pairs surges by 43\% and 50\% by using RE+PV and RE+LV respectively, which is a remarkable improvement. Considering the number of neg-pos pairs, our proposed approach is also superior to random selection and ELO-DCG \cite{long2010active,long2014active}. Generally, the number of neg-pos pairs is related to the percentage of the irrelevant webpages in top $K$ since using more neg-pos pairs to train an LTR model, it would be easier to distinguish relevant webpages from irrelevant webpages. Using RE+PV the number of neg-pos pairs dramatically raises by 50\% compared with random selection and it is also boosted by 34\% compared to the existing active learning approach -- ELO-DCG \cite{long2010active,long2014active}.

\paragraph{LTR Performance Comparisons} In this experiment, we use the valid query-webpage pairs obtained by various strategies to train LTR models (GBRank models with cross-entropy loss) and compare the ranking quality of these LTR models on our validation dataset of 1,000 queries. The ranking quality is measured using DCG. Fig. \ref{dcg-offline} shows the comparison among different approaches. 

Let's first pay attention to base100 -- the top two sub-figures in Fig. \ref{dcg-offline}. The proposed approach RE+PV achieves better performance than random selection and ELO-DCG \cite{long2010active,long2014active}. We can see that using RE+PV selected queries to train GBRanK, $DCG_4$ raises faster than its counterparts, such as random selection, ELO-DCG and RE. Compared with random selection, the relative improvement of $DCG_4$ ranges from 0.35\% to 1.38\% using RE+PV. And compared to the existing work ELO-DCG \cite{long2010active,long2014active}, the proposed RE+PV boosts $DCG_4$ by at least 0.37\%. While random selection outperforms ELO-DCG and RE when selecting more training data. The possible reason is that ELO-DCG and RE are biased by query frequency, \eg, ELO-DCG prefers high-frequency queries, whereas RE is in favor of low-frequency queries. Interestingly, ELO-DCG and RE perform similarly to each other, though the distributions of the selected queries are different. Looking at PV and LV that are related to the diversity among webpages, both outperform random selection in most cycles. When it comes to $R_{01}$ (top-right sub-figure in Fig. \ref{dcg-offline}), the proposed RE+PV also achieves competitive performance compared with its counterparts. \Eg, $R_{01}$ drops by at most 6.31\% compared to random selection. Although ELO-DCG is able to obtain higher $DCG_4$ scores at the beginning of AL cycles, it performs even worse considering the metric $R_{01}$. The reason is that ELO-DCG selects more webpages with label 2 but fewer webpages with label 0, hence, it is difficult for GBRank to distinguish irrelevant webpages. 

Moving on to base500 (bottom sub-figures in Fig. \ref{dcg-offline}), our proposed approach -- RE+PV also achieves higher $DCG_4$ than random selection and ELO-DCG \cite{long2010active,long2014active}. The relative improvement of $DCG_4$ is at most 0.84\% compared to random selection. Moreover, $R_{01}$ shows decreases in most cycles, \eg, in cycle 9, $R_{01}$ drops by 3.76\% using RE+PV. 

We also present the relative improvement of the average $DCG_4$ over AL cycles for each category in Table \ref{avg_dcg}. We can easily find that most AL approaches obtain better performance on high-frequency queries compared with random selection, \eg, for category 0, using RE, RE+0.5PV and RE+LV boost $DCG_4$ by 1.13\% compared with random selection in the scenario of base100. $DCG_4$ also increases by 0.42\% using ELO-DCG \cite{long2010active,long2014active} for category 0 since ELO-DCG selects more high-frequency queries (see Fig.\ref{bucket_distribution}). By contrast, for low-frequency queries, the performances of different approaches vary, \eg, 
$DCG_4$ for category 9 obtained by ELO-DCG drops by 0.74\% compared with random selection, whereas $DCG_4$ obtained by our proposed approach -- RE+PV is 1.23\% higher than using random selection. Interestingly, the performances of using $bs=100$ and $bs=500$ in category 3 are different. We can see that the test models except for PV and RE outperform random selection when using $bs=100$, while they are inferior to random selection when $bs=500$. The possible reason is that for the queries belonging to category 3, it is difficult to annotate the relevance scores \footnote{For the queries belonging to category 0, it is easy to annotate since many webpages are highly relevant to the queries. Likewise, we can easily annotate queries belonging to category 9, since the webpages are irrelevant to queries.}, resulting in many noisy labels, and selecting more queries introduces more noisy labels, hence the performance reduces when using $bs=500$.

Note that we also conduct ablation studies on the hyper-parameter $\alpha$ in Eq. (\ref{acquisition}), finding that $\alpha=1$ is a better choice and in our online experiments we use this value.

\subsection{Online Experiments and Results}
To report the online performance of our proposals, we carried out an A/B Test with real-world web traffics. Note that online testing is expensive and time-consuming, so we only compare our proposed model with the existing baseline that employs random selection in Baidu search engine.

\subsubsection{A/B Test Setups}

\begin{figure}[t]
    \centering
    \includegraphics[width=0.49\linewidth]{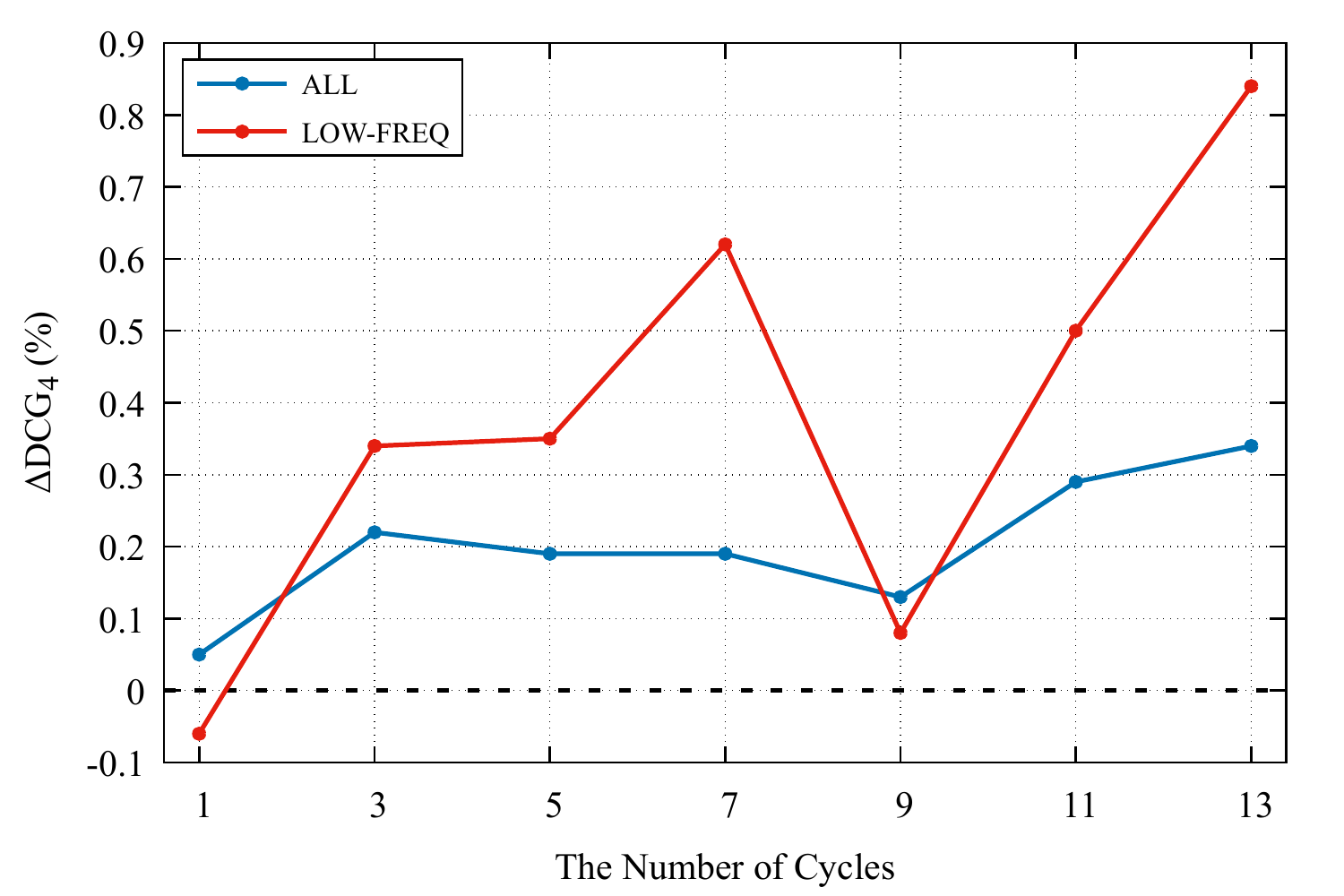}
    \hfill
    \includegraphics[width=0.49\linewidth]{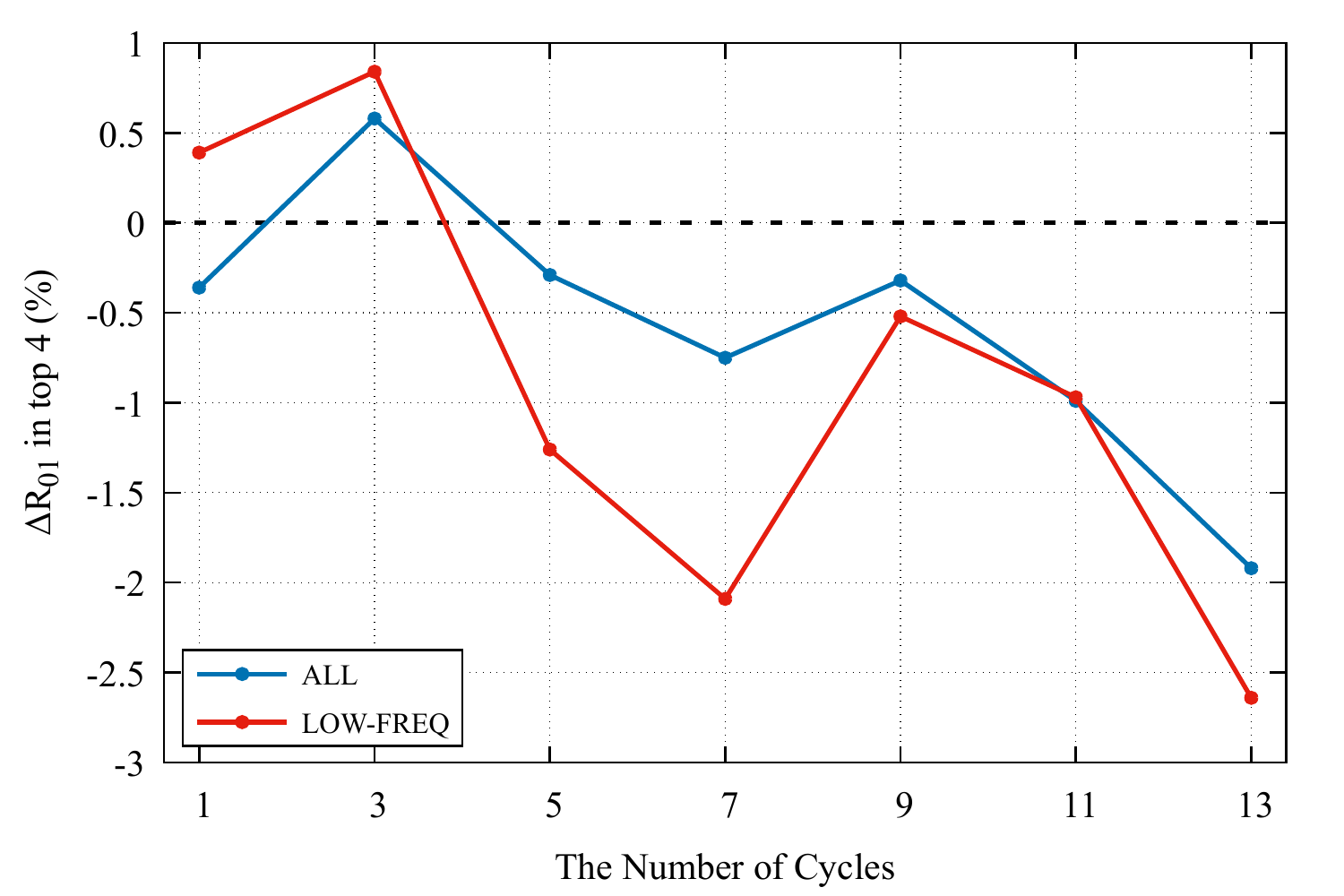}
    \caption{The online performance. We only report the relative improvement with p-value $<$ 0.05 over the baseline. Left: the performance based on $DCG_4$. Right: the performance based on $R_{01}$.}
    \label{dcg-online}
\end{figure}

We used 0.6\% real-world web traffics on Baidu search engine to conduct the A/B test, where the 0.6\% traffics have been randomly partitioned into two folders (0.3\%) each to evaluate the performance of our proposals (RE+PV) and random selection respectively. This online A/B test lasted for 13 days/cycle.
In each cycle, we use our proposed approach to select 500 queries from the historical query pool composed of hundreds of millions of queries and each query has 100 retrieved webpages. Then we filter out the pornographic webpages and the webpages forbidden by the government, resulting in around 60 webpages for each query. After that, we hire people to annotate the relevance scores and each query-webpage pair has at least 6 scores annotated by different workers. Finally, our expert annotator will evaluate the quality of annotations to make sure that the accuracy is higher than 85\% and then we use the weighted sum of the scores as the final label to train our LTR model. Similar to the offline experiments, we also calculate $\Delta DCG_4$ and $\Delta R_{01}$ between the two methods based on the ground-truth annotation results.

As we have mentioned, we can only label 500 queries per day, hence, in the online experiments we use the $bs=500$. Also, note that our query pool is composed of hundreds of millions of historical queries, therefore, the selection criteria should be as simple as possible, otherwise, our cluster cannot handle the selection in a few hours. Moreover, since we need to update the LTR model every day, and to avoid some unexpected influences on the search engine, we just use the 0.6\% traffic to test the proposed approach. 


%

\subsubsection{Online Performance}

The comparison is shown in Fig. \ref{dcg-online}, where we only present the relative improvements. Compared to random selection, the proposed RE+PV is able to boost $DCG_4$ in all cycles and the largest relative improvement is around 0.35\% when considering all queries. We also compare RE+PV to random selection on low-frequency queries (categories 7, 8, 9) and $DCG_4$ increases by at most 0.85\%, indicating that the proposed approach benefit low-frequency query search. In terms of $R_{01}$, our proposed approach -- RE+PV can also reduce the percentage of irrelevant webpages in the top $K$ results, \eg, $R_{01}$ decreases by at most 1.9\% considering all queries, while it drops by at most 2.6\% on low-frequency queries. Basically, the online performance is consistent with our offline results and the proposed active learning approach outperforms the baseline.

\section{Conclusion and Future Work}

In this work, we revisited the problem of active learning for ranking webpages in the context of Baidu search, where the key problem is to establish the training datasets for learning to rank (LTR) models. Given trillions of queries and relevant webpages retrieved for every query, the goal of active learning is to select a batch of queries for labeling and trains the current LTR model with the newly-labeled datasets incrementally, where the labels here refer to the ranking score of every webpage under the query. To achieve the goals, for every query, this work proposed two new criteria--\emph{Ranking Entropy (RE)} and \emph{Prediction Variances (PV)}  that could measure the \emph{uncertainty of the current LTR model to rank webpages in a query} and the \emph{diversity of ranking scores for webpages in a query} respectively. Specifically, RE estimates the entropy of relevant webpages under a query produced by a sequence of online LTR models updated by different checkpoints, using a Query-By-Committee (QBC) method, while PV estimates the variance of prediction results for all relevant webpages under a query. Our experimental observations find that RE may pop low-frequency queries from the pool for labeling while PV prioritizes high-frequency queries more. Further, the estimate of PV significantly correlates to the diversity of true ranking scores of webpages (annotated by humans) under a query and correlates to the information gain of LTR. Finally, we combine these two complementary criteria as the sample selection strategies for active learning. Extensive experiments with comparisons to baseline algorithms show that the proposed approach could train LTR models achieving higher DCG (\ie, $\Delta$DCG$_4$=0.35\%$\sim$1.38\% in offline experiments, $\Delta$DCG$_4$=0.05\%$\sim$0.35\% in online experiment) using the same budgeted labeling efforts, while the proposed strategies could discover 43\% more valid training pairs for effective training. \revisionsec{Note that the queries selected by active learning are more informative and we can use fewer queries to train LTR models, achieving satisfying performance, which saves millions of yuan per year.}

\revision{Recently, deep learning approaches have been applied to web search and real-world products, for example, in Baidu search engine, we have used pre-trained big models \cite{zou2022pre}, Kaleido-BERT \cite{zhuge2021kaleido} and AliCoCo \cite{luo2020alicoco} employs big models for E-commerce search and both of them introduce specific knowledge into the search engine. One future direction should be cold-started active learning, \ie, using the pre-trained models to select samples for annotation. In the field of web search, artificial intelligence-generated content (AIGC) should draw much attention. ChatGPT \cite{ouyang2022training} has shown the strong ability of big models for question answering. Also, some works on text-to-image and image-to-text translation \cite{wang2019describing,wang2020neighbours,wang2020compare,zhang2021ernie,wang2022diversity,wang2022distinctive} should provide content for search engines, therefore, another future direction could be combining both AIGC and current search engines to well satisfy users' demands.}

\bibliographystyle{unsrt}
\bibliography{ref}

\end{document}